\documentclass[]{spie}  

\usepackage[square,numbers,sort&compress,comma]{natbib}
\usepackage{amsmath,amsfonts,amssymb}
\usepackage{graphicx}
\usepackage[export]{adjustbox}
\usepackage{caption}
\graphicspath{{./}{./figures/}}
\usepackage[colorlinks=true, allcolors=blue]{hyperref}
\usepackage{threeparttable}
\usepackage[usenames,dvipsnames,svgnames]{xcolor}
\usepackage{multirow}

\newcommand{\eq}{\begin{equation*}}
\newcommand{\qe}{\end{equation*}}
\usepackage{float}
\graphicspath{{figures/}{figures/hardware/}{figures/noise/}{figures/observation/}{figures/spectra/}}

\usepackage{aas_macros}
\usepackage[]{natbib}
\usepackage{amsmath}
\usepackage{amssymb}
\usepackage{xspace}
\usepackage{xifthen}
\usepackage{eso-pic}

\mathchardef\mhyphen="2D

\newlength{\dhatheight}

\newcommand{\unit}[1]{\ensuremath{\mathrm{\,#1}}\xspace}

\newcommand{\rmspix}{\unit{rms/pix}}

\title{Astronomical Spectroscopy with Skipper CCDs: First Results from a Skipper CCD Focal Plane Prototype at SIFS 
}

\author[1,2,*]{Edgar {Marrufo Villalpando}}
\author[3,2,4,$\dagger$]{Alex Drlica-Wagner}
\author[2]{Brandon Roach}
\author[5]{Marco Bonati}
\author[3]{Abhishek Bakshi}
\author[6]{Julia Campa}
\author[3]{Gustavo Cancelo}
\author[5]{Braulio Cancino}
\author[3]{Claudio R. Chavez}
\author[12]{Fernando Chierchie}
\author[3]{Juan Estrada}
\author[3]{Guillermo {Fernandez Moroni}}
\author[7]{Luciano Fraga}
\author[8]{Manuel E.\ Gaido}
\author[9]{Stephen E.\ Holland}
\author[1]{Rachel Hur}
\author[3]{Michelle Jonas}
\author[5]{Peter Moore}
\author[12]{Eduardo Paolini}
\author[10, 11]{Andrés A.\ {Plazas Malagón}}
\author[3]{Leandro Stefanazzi}
\author[3]{Javier Tiffenberg}
\author[3]{Ken Treptou}
\author[3]{Sho Uemura}
\author[3]{Neal Wilcer}

\affil[1]{\small Department of Physics, University of Chicago, Chicago, IL 60637, USA}
\affil[2]{\small Kavli Institute of Cosmological Physics, University of Chicago, Chicago, IL 60637, USA}
\affil[3]{\small Fermi National Accelerator Laboratory, Batavia, IL 60510, USA}
\affil[4]{\small Department of Astronomy \& Astrophysics, University of Chicago, Chicago, IL 60637, USA}
\affil[5]{\small Cerro Tololo Inter-American Observatory, NSF’s National Optical-Infrared Astronomy Research Laboratory, Casilla 603, La Serena, Chile}
\affil[6]{\small Departamento de Física, Universidad de Córdoba, Córdoba, España}
\affil[7]{\small Laborat\'{o}rio Nacional de Astrof\'{i}sica LNA/MCTI, 37504-364, Itajub\'{a}, MG, Brazil}
\affil[8]{\small Universidad de Buenos Aires, Facultad de Ciencias Exactas y Naturales, Departamento de Física. Buenos Aires, Argentina}
\affil[9]{\small Lawrence Berkeley National Laboratory, One Cyclotron Rd, Berkeley, CA 94720, USA}
\affil[10]{\small Kavli Institute for Particle Astrophysics and Cosmology, Stanford University, Stanford, California, USA}
\affil[11]{\small SLAC National Accelerator Laboratory, Menlo Park, California, USA}

\affil[12]{\small Instituto de Investigaciones en Ingeniería Eléctrica “Alfredo C. Desages” CONICET,
Bahía Blanca, Argentina}

\authorinfo{$^*$~emarrufo@uchicago.edu \\ $^\dagger$~kadrlica@fnal.gov}

\begin{document} 
\maketitle

\begin{abstract}

We present the first on-sky results from an ultra-low-readout-noise Skipper CCD focal plane prototype for the SOAR Integral Field Spectrograph (SIFS). The Skipper CCD focal plane consists of four 6k $\times$ 1k, 15 $\mu$m pixel, fully-depleted, $p$-channel devices that have been thinned to $\sim$250 $\mu$m, backside processed, and treated with an anti-reflective coating. These Skipper CCDs were configured for astronomical spectroscopy, i.e., single-sample readout noise $< 4.3$\,e$^-$\,rms/pixel, the ability to achieve multi-sample readout noise $\ll 1\,$e$^-$\,rms/pixel, full-well capacities $\sim$40,000--65,000 e$^-$, low dark current and charge transfer inefficiency ($\sim 2$ $\times$ $10^{-4}$\,e$^-$/pixel/s and 3.44 $\times$ $10^{-7}$, respectively), and an absolute quantum efficiency of $\gtrsim 80\%$ between 450\,nm and 980\,nm ($\gtrsim 90\%$ between 600\,nm and 900\,nm). We optimized the readout sequence timing to achieve sub-electron noise ($\sim 0.5$\,e$^-$\,rms/pixel) in a region of 2k $\times$ 4k pixels and photon-counting noise ($\sim 0.22$\,e$^-$\,rms/pixel) in a region of 220 $\times$ 4k pixels, each with a readout time of $\lesssim 17$ min. We observed two quasars (HB89\,1159$+$123 and QSO\,J1621$–$0042) at redshift $z \sim 3.5$, two high-redshift galaxy clusters (CL\,J1001$+$0220 and SPT-CL\,J2040$-$4451), an emission line galaxy at $z = 0.3239$, a candidate member star of the Bo\"{o}tes~II ultra-faint dwarf galaxy, and five CALSPEC spectrophotometric standard stars (HD074000, HD60753, HD106252, HD101452, HD200654). We present charge-quantized, photon-counting observations of the quasar HB89\,1159$+$123 and show the detector sensitivity increase for faint spectral features. We demonstrate signal-to-noise performance improvements for SIFS  observations in the low-background, readout-noise-dominated regime. We outline scientific studies that will leverage the SIFS-Skipper CCD data and new detector architectures that utilize the Skipper floating gate amplifier with faster readout times.
    
\end{abstract}


\section{INTRODUCTION}
\label{sec:intro}  

The invention of silicon charge-coupled devices (CCDs) \citep{Boyle:1970} led to a revolution in our understanding of the Universe. The exceptional sensitivity, stability, and precision of CCDs has led to fundamental discoveries about the growth and content of the Universe, the formation of stars and galaxies, and planets around stars other than our Sun. In the coming decades, CCDs will be used to address some of the biggest questions in science: What are the dark matter and dark energy that comprise most of our Universe? How do galaxies form? Are there Earth-like planets that could host life around other stars?

One of the main technical limitations for answering these questions comes from the electronic noise that is introduced into astronomical observations during the read out of CCD detectors.
It is only recently that novel Skipper-CCD technology has allowed CCDs to achieve noise levels that are much lower than the output signal produced by a single photo-electron \cite{Tiffenberg:2017}.
Here, we present the first on-sky observations with astronomy-grade, ultra-low-noise, photon-counting Skipper CCDs using the Southern Astrophysical Research (SOAR) Telescope Integral Field Spectrograph (SIFS) \cite{10.1117/12.461977, 10.1117/12.857698}.  We highlight improvements in sensitivity from the ultra-low-noise capabilities of the Skipper CCD and offer an outlook on future detectors implementing the Skipper floating gate amplifier for cosmological measurements. To our knowledge, this is the first on-sky demonstration using Skipper CCDs for astronomical spectroscopy.            

This manuscript describes key aspects of the Skipper CCD detector characterization, focal plane assembly, readout configuration, observation planning, and first on-sky demonstration of Skipper CCDs. 
A companion paper (Bonati et al., 2024 \cite{Bonati:2024}) focuses on the electrical, mechanical, and cryogenic engineering work performed at the observatory to enable this demonstration.
In Section \ref{sec:sifs_skipper}, we summarize detector characterization and optimization results, which confirm that Skipper CCDs perform similarly to other thick, fully-depleted $p$-channel CCDs (e.g., stability, linear response, large dynamic range, and high quantum efficiency). We also present characterization measurements enabled by photon-counting and directed toward studying the detector performance at low signal levels (e.g., low-signal non-linearities). Section \ref{sec:comissioning} summarizes the installation and commissioning of the Skipper CCD focal plane. In Section \ref{sec:obs}, we outline the Region of Interest (ROI) \cite{Chierchie_2021,10.1117/12.2562403} observation strategy developed to achieve sub-electron readout noise and photon-counting while limiting readout times to $<20$ min.
This section also describes the astronomical targets chosen to demonstrate the signal-to-noise improvements enabled by the Skipper's ultra-low noise. Section \ref{sec:results} presents results from the observations, demonstrating signal-to-noise improvements and describing limitations of the as-built system. In Section \ref{sec:sci}, we present two future science cases: the study of decaying axion-like-particle (ALP) dark matter and the study of a candidate member star of an ultra-faint dwarf (UFD) galaxy. These two scientific applications are intended to leverage the low-noise capabilities of Skipper CCDs. We conclude in Section \ref{sec:end} by summarizing our results and offering an outlook on future implementations of the Skipper amplifier with improved readout times.

\section{SIFS Skipper CCDs}
\label{sec:sifs_skipper}  

Ultra-low-noise Skipper CCDs have been identified as a detector technology for astronomical applications in the readout-noise-dominated, low-background regime (e.g., astronomical spectroscopy of faint sources) \cite{Tiffenberg:2017, 10.1117/12.2562403}. In Drlica-Wagner et al. (2020) \cite{10.1117/12.2562403}, we performed the first optical characterizations of a Skipper CCD and showed that the anti-reflective (AR) coated, backside illuminated, 250 $\mu$m thick, detector could achieve a relative quantum efficiency (QE) $> 75 \%$ from 450 nm to 900 nm and a full-well capacity $\sim 34{,}000$ e$^-$ with no voltage optimizations. These results motivated further work to optimize Skipper CCDs for astronomy. In Villalpando et al. (2022) \cite{10.1117/12.2629475}, we outlined plans to assemble four 6k $\times$ 1k Skipper CCDs in a focal plane array  as a prototype to replace the original 4k $\times$ 4k CCD231-84 e2v detector and conduct the first astronomical measurements with Skipper CCDs at SIFS. Critically, the Skipper CCD focal plane prototype is installed in an identical SOAR Dewar and maintains the optical alignment of the original SIFS system. In addition, the SDSU-III controller is replaced by four Low Threshold Acquisition (LTA) boards \cite{10.1117/1.JATIS.7.1.015001} to read the 16 output channels (4 channels per Skipper CCD). In Villalpando et al. (2024) \cite{Villalpando_2024}, we report results from the characterization and optimization of the 6k $\times$ 1k, 15 $\mu$m pixels, 250 $\mu$m thick, fully-depleted Skipper CCDs used at SIFS, which we summarize below.   

\begin{figure} [t]
   \begin{center}
   \begin{tabular}{l l} 
   \includegraphics[height=6.5cm]{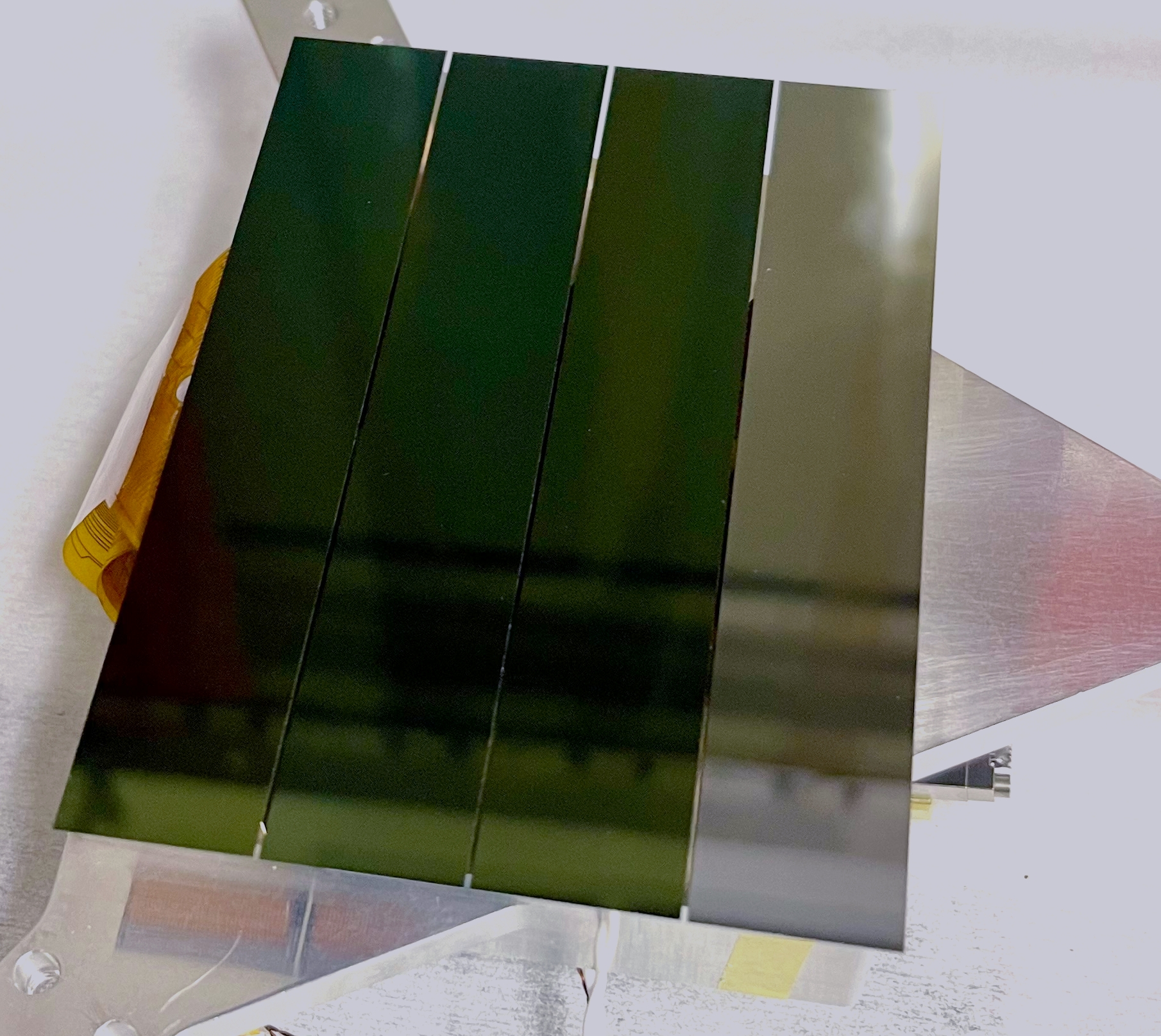} 
   \includegraphics[height=6.5cm]{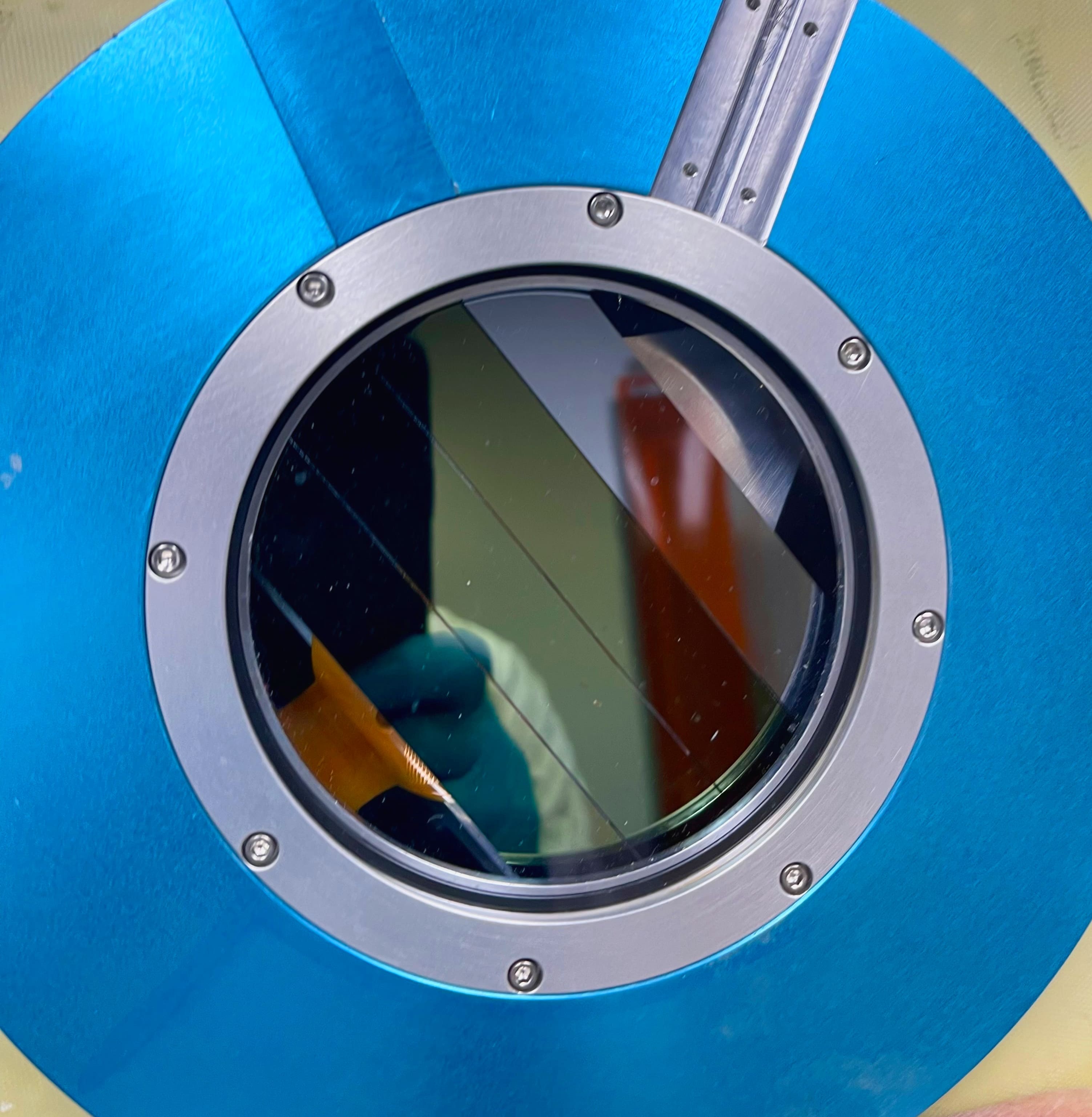}    
   \end{tabular}
   \end{center}
   \caption{\label{fig:focal_plane} 
   \textbf{Left}: Skipper CCD focal plane prototype for SIFS. Four astronomy-grade Skipper CCDs are placed on a custom mount with three arms that attach to a standard SOAR Dewar. The focal plane spans an area of 6k $\times$ 4k pixels. The four devices are separated by gaps of $\sim$400\,$\mu$m.  \textbf{Right}: Focal plane inside the standard SOAR Dewar with a fused-silica window for exposure to light.}
\end{figure}

Eight backside illuminated, astronomy-grade Skipper CCDs (AstroSkippers) were fabricated for the SIFS focal plane prototype. These devices were designed at
Lawrence Berkeley National Laboratory (LBNL), fabricated at Teledyne DALSA, and packaged at Fermi National Accelerator Laboratory (Fermilab). The AstroSkippers are conventional $p$-channel Skipper CCDs, fabricated on high resistivity ($>$ 5 k$\Omega$cm), $n$-type silicon. They were thinned from the standard wafer thickness of 650-675 $\mu$m to 250 $\mu$m at a commercial vendor and then backside processed at the LBNL Microsystems Laboratory. Backside processing and AR coating follow the same procedures as used for the Dark Energy Spectroscopic Instrument (DESI) devices, resulting in high QE from the near-infrared (NIR) to the near-ultra-violet (NUV) \cite{Bebek:2017,Villalpando_2024}. We use four AstroSkippers for the construction of the focal plane prototype (Figure \ref{fig:focal_plane}).  

\begin{table}[t]
\centering
\caption{\label{tab:astroskipper}
AstroSkipper Key Performance Metrics
}
\begin{tabular}{l c c c}
\hline
Characteristic  & Value  & Unit \\
\hline \hline
Readout Speed & 25 & kpixels/s \\
Single-Sample Readout Noise & $<$ 5 & e$^-$rms/pixel  \\
Multi-Sample Readout Noise & 0.18 & e$^-$ rms/pixel  \\
Dark Current & $2 \times 10^{-4}$ & e$^-$/pixel/s  \\
Full-Well capacity & $\sim 40,000 - 60,000$ &e$^-$  \\
Non-Linearity & $<$ 0.05$\%$ and $<1.5 \%$ (low signal) & ...  \\
\multirow{ 2}{*}{Absolute Quantum Efficiency} & $\gtrsim 80 \% $ (450 nm to 980 nm) & \multirow{2}{*}{...} \\
 & $\gtrsim 90\% $ (600 nm to 900 nm) &   \\
Charge Transfer Inefficiency & $3.44 \times 10^{-7}$ & ...  \\
\hline
\vspace{1em}
\end{tabular}
\end{table}

\subsection{Key Performance Metrics for the AstroSkipper}
We summarize key results from the characterization and optimization of the Skipper CCDs used at SIFS \cite{Villalpando_2024}. Key performance metrics for these devices are collected in Table \ref{tab:astroskipper}.

\noindent \textit{Readout Time:} The pixel integration time window, which includes the pedestal and signal components, was optimized as a function of the readout noise. The integration window optimization reduced the total pixel readout time from $\sim 200$ $\mu$s/pixel to $\sim 40$ $\mu$s/pixel (i.e., 25 kpixels/s) for a single amplifier and pixel sample. Laboratory tests demonstrated single-sample readout noise of $<$ 4.3\,e$^-$\,\rmspix for all detectors' amplifiers with the fast readout time configuration. We define multi-sample ROIs \cite{Chierchie_2021,10.1117/12.2562403} to achieve sub-electron readout noise in a subset of detector pixels in a readout time of $<20$ min. This is a significant improvement over the unoptimized readout time configuration, which would take $\sim 50$ min to achieve 0.5\,e$^-$\,rms/pixel. We detail the ROIs used for a set of astronomical observations with SIFS in Section \ref{sec:obs}.   

\noindent \textit{Non-linearity:} Low-signal non-linearities (a few tens of electrons) are poorly characterized in conventional scientific CCDs since these lack the precision to measure charge with electron resolution \cite{Bernstein_2017}. In contrast, non-linearities can be well-characterized for all electron occupancies in Skipper CCDs. For the SIFS application, we target signal-starved observations where the expected signal rate per exposure is $\lesssim$ 15 e$^-$. Therefore, understanding the performance of the Skipper CCD at low signals was an essential part of the lab testing. In the photon-counting regime, we probe non-linerities by measuring variations in the relationship between the number of electrons in each pixel and the signal readout value in analog-to-digital units (ADUs). Non-linearity measurements are performed from electron peak-to-peak separation, i.e., we fit each electron peak with a Gaussian and compute the gain from each peak by dividing the mean value of the difference between the electron peak and the 0 e$^-$ peak by the peak's assigned electron number through counting. We represent non-linearity as the deviation from unity of the ratio of the gain computed from each electron peak to the fixed, conventional gain measured from the photon transfer curve (PTC). We measure low-signal non-linearities $< 1.5 \%$ for 0 to 50 e$^-$ and $< 0.05 \%$ for high signal levels ($>1500$ e$^-$) where we implemented a conventional least-squares fitting method for measuring non-linearity in scientific CCDs \cite{10.1117/12.2559203, Bonati:2024}.  

\noindent \textit{Voltage Optimization:} An important result from the optimization of the AstroSkipper CCDs was the trade-off between full-well capacity and clock induced charge (CIC) \cite{Janesick:2001}. When operating Skipper CCDs with a low voltage configuration (i.e., voltages used for dark matter searches \cite{Barak:2020}), the CIC rate is $\sim 1.45 \times 10^{-3}$ e$^-$/pixel/frame with a full-well capacity of $\sim$ 900 e$^-$. However, for the SIFS application where calibration data products have signals $>40{,}000$ e$^-$, we optimized clock voltages to accommodate the expected signals from SIFS while maintaining the lowest CIC rate possible. For the observations, we implemented two voltage configurations based on the expected signals, i.e., for calibration data products we operated with a full-well capacity $> 40{,}000$ e$^-$ and a CIC rate of $\sim$3 e$^-$/pixel/frame, while for science data products we operated with a full-well capacity of ${\sim} 10{,}000$ e$^-$  and a CIC of $<$1 e$^-$/pixel/frame. These CIC rates are subdominant compared to the external contributions to the dark rate  in the science exposures (e.g. external light leaks in a $>600$\,s exposure). We note that CIC is an important issue that might limit signal-to-noise for observation where background contributions to the noise are low and the expected signals are in  the few electron regime. However, shaping the clock pulse rise time and sharpness, which play a critical role in CIC generation, has been shown to reduce CIC in electron multiplying CCDs (EMCCDs) \cite{10.1117/12.2055346, 10.1117/12.2232879}.

\noindent \textit{Absolute Quantum Efficiency:} We performed the first absolute QE measurements for astronomy-grade Skipper CCDs and demonstrate better performance than previously reported \cite{10.1117/12.2562403}. The reliability of the absolute QE measurement depends on an accurate determination of the incident power at the location of the CCD housed inside the lab testing Dewar. To derive robust QE measurements, we employed calibrated silicon photodiodes mounted on the external integrating sphere and on an AstroSkipper package mounted inside the Dewar. Multiple calibration factors, i.e., the ratio of the power measurements at both photodiodes, were taken into account when estimating uncertainties in the measurement. We find QE $\gtrsim 80 \%$ from 450\,nm to 980\,nm, and QE $\gtrsim 90 \%$ from 600\,nm to 900\,nm for all tested AstroSkippers with an uncertainty of $< 6\%$ for all wavelengths \cite{Villalpando_2024}.         

\section{INSTALLATION AND COMMISSIONING}
\label{sec:comissioning}  

Installation and commissioning of the SIFS Skipper CCD focal plane prototype took place March 25--29, 2024 in advance of the first on-sky observations on March 31, 2024. Details on the installation of the Skipper CCD focal plane and final engineering performance results can be found in Bonati et al. (2024) \cite{Bonati:2024}, published in this volume. Here, we highlight important considerations regarding the performance of the Skipper CCD focal plane. The four devices are separated by gaps of $\sim$400\,$\mu$m, which lead to gaps of $\sim$275\,\AA\ in the wavelength coverage \cite{Bonati:2024}. The Dewar installs into a kinetic mount, which provides optical alignment with respect to the light beam. We measured the full width at half maximum of the fiber traces  from a series of masked flat-fields taken with a quartz lamp across the entire field of view and determined that variations were around $13 \%$, which was comparable to the original SIFS alignment so no re-alignment was needed. Laboratory tests of the fully assembled Skipper CCD array showed correlated noise ($\sim 17$\,e$^-$\,rms/pixel) on two detectors, despite symmetry in the readout electronics outside the Dewar. We found that reading multiple-samples per pixel decreased the noise faster than expected from uncorrelated Gaussian noise (i.e., $ \sigma_N < \sigma_1/\sqrt{N_{\rm samp}}$). After $\sim 12$ samples per pixel, we recover Gaussian noise behavior (see Section \ref{sec:results}). During commissioning, it was noted that stray light in the SIFS enclosure was contributing a shutter-open dark rate of $\mathcal{O}(10^{-1})$ e$^-$/pixel/s. This stray light affected observations taken on March 31, 2024, but it was reduced by an order of magnitude for observations on April 9, 2024 (Section~\ref{sec:results}). 

\begin{figure}[t!]
    \centering
    \includegraphics[width=0.90\textwidth]{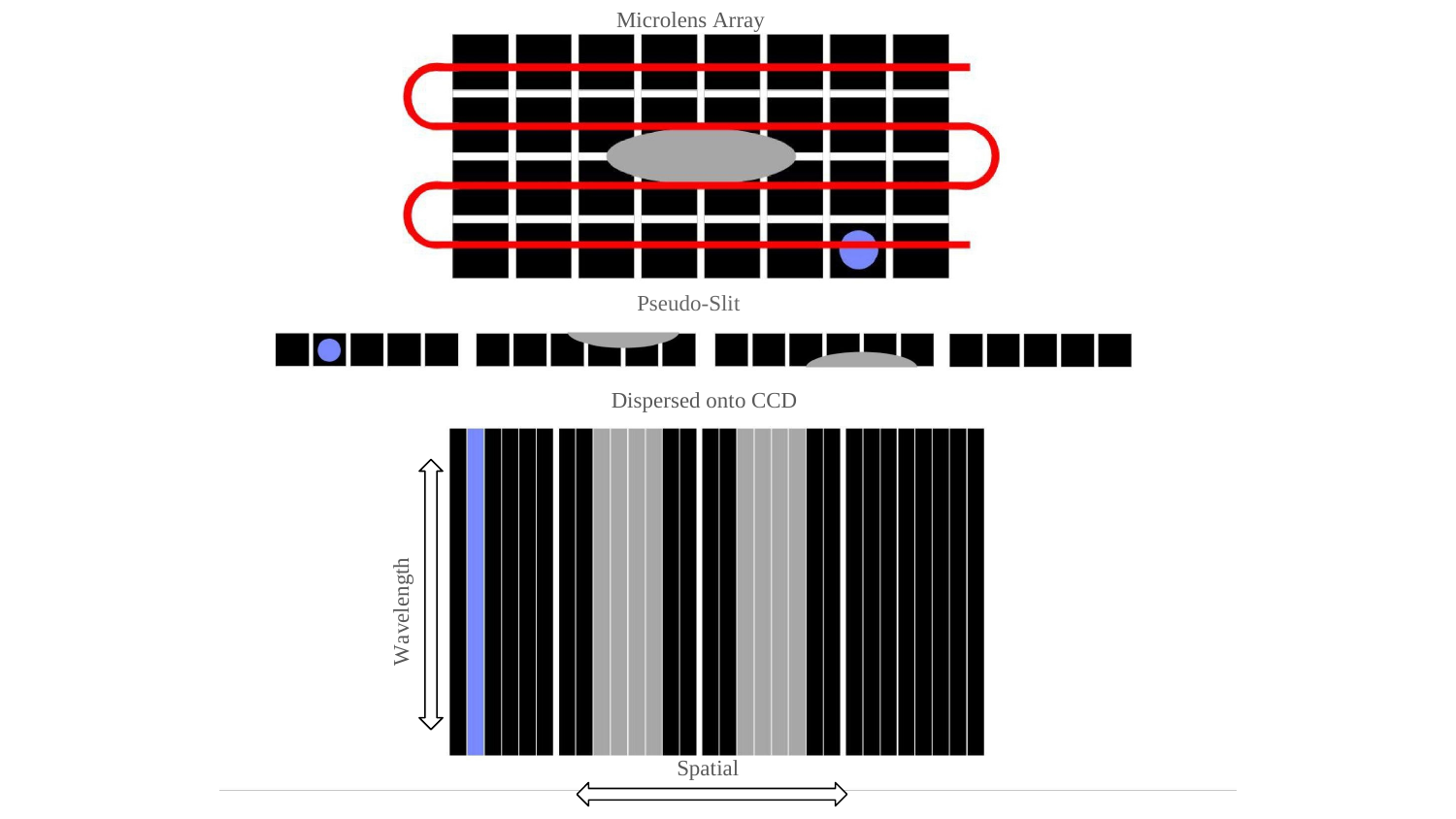}
    \vspace{1em}
    \caption{Mapping of the signal from the SIFS optics to the Skipper CCD focal plane. The microlens array collects light from the astronomical target and channels the light into the fibers, which are arranged into a pseudo-slit following a serpentine mapping (red curved line). The fibers disperse the light onto the focal plane; the wavelength direction runs along the Skipper CCDs' rows and the fibers' spatial position runs along the detectors' columns.}
    \label{fig:sifs_skipper_map}
\end{figure}

\section{OBSERVATION STRATEGY}
\label{sec:obs}

Science-quality observations were acquired with the SIFS Skipper CCD focal plane on March 31 and April 9, 2024 during time allocated to Brazilian collaborators on SOAR.
The science verification program was intended to demonstrate the performance of Skipper CCDs on the sky for the first time. The observational parameters were chosen to showcase the signal-to-noise improvement possible from the reduced readout noise of the Skipper CCD in the low-signal regime. Therefore, we chose astronomical targets that satisfy our science case (see Section \ref{sec:sci}) and prioritized reducing readout noise over a detector region-of-interest in order to minimize readout times.
In several cases, we reduced integration times to several minutes in order to artificially reduce the signal strength and place the observations in the readout noise dominated regime.

SIFS is equipped with a 1300-element Integral Field Unit (IFU) that operates through two main components: the fore-optics and the bench spectrograph. The fore-optics consists of a 26 $\times$ 50 array of 1mm $\times$ 1mm square microlenses, which channel light into the fibers. The bench spectrograph receives these aligned fibers arranged into a pseudo-slit. A set of Volume Phase-Holographic transmission gratings ensure the system provides the desired resolution and wavelength coverage, adjustable from $5000 \lesssim R \lesssim 20000$ and $3500 \lesssim \lambda \lesssim 10000$\,Å, respectively. The IFU has a fiber scale of 0.30 arcsec/fiber which produces a field of view of 15 $\times$ 7.8 arsec$^2$ \cite{10.1117/12.461977,10.1117/12.857698}. Figure \ref{fig:sifs_skipper_map} shows the spatial and wavelength mapping from the microlens array to the Skipper CCD focal plane. The gap between detectors causes $\sim$2\% of the wavelength coverage to be lost. Therefore, object targeting becomes important to ensure that important features are mapped to an active region on the focal plane. 

\begin{figure}[t!]
    \centering
    \includegraphics[width=0.96\textwidth]{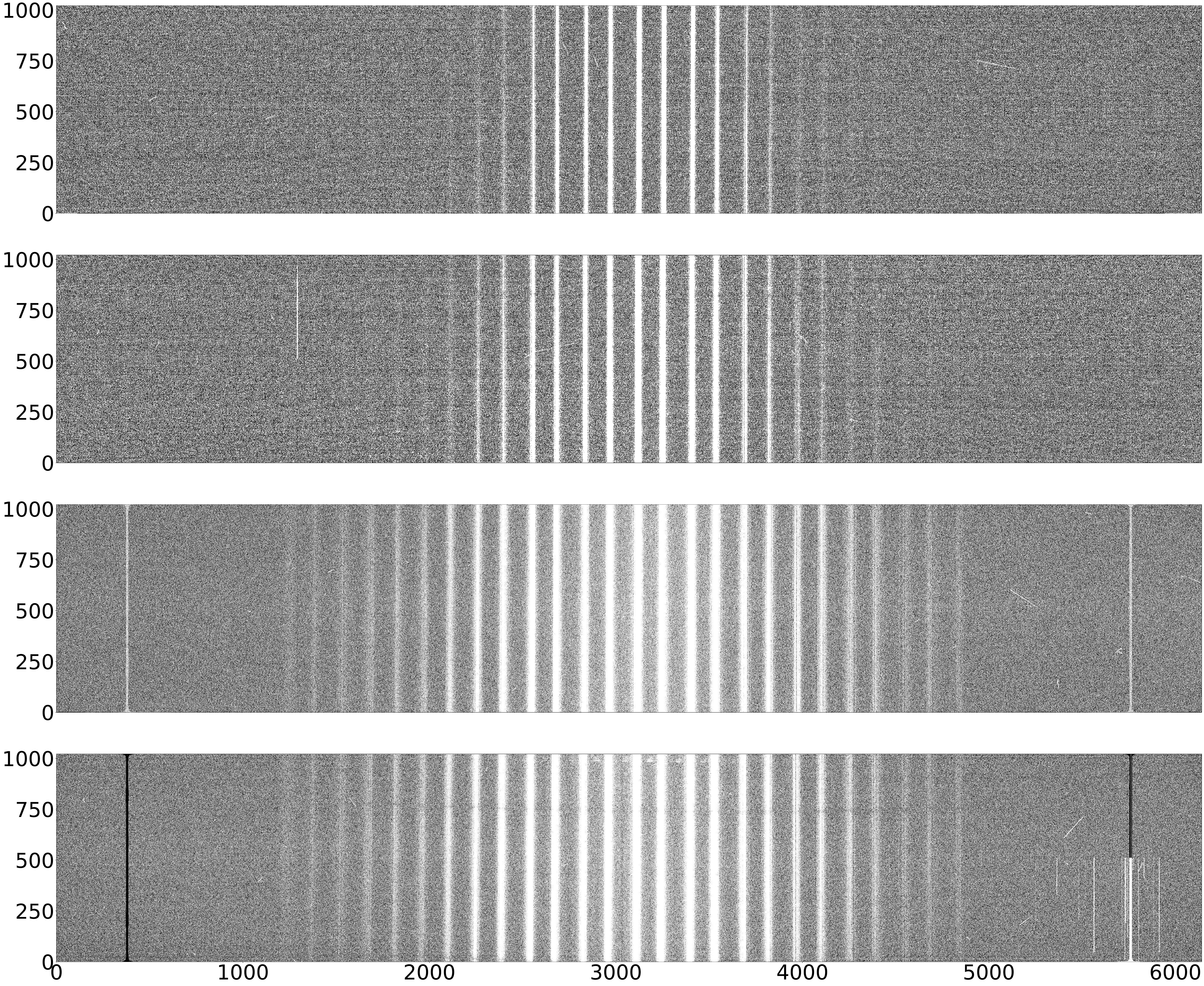}

    \vspace{1em}
    \caption{Skipper CCD focal plane array data frames. The four frames show the full $6$k $\times$ 4k physical pixels of the four Skipper CCDs. Each frame is composed of the four amplifiers in each detector. We show spectroscopic data from a CALSPEC \cite{Bohlin_2014} standard calibration star (HD101452); the fiber bundles span the 4k $\times$ 4k SIFS FoV. There are ``hot''  columns present outside the 4k $\times$ 4k FoV, which do not affect data reduction. The gaps between the Skipper CCDs are ${\sim} 400$\,$\mu$m (${\sim}27$ pixels) wide and are not shown to scale.}
    \label{fig:FOV}
\end{figure}


\begin{figure}[t!]
    \centering
     \includegraphics[width=0.99\textwidth]{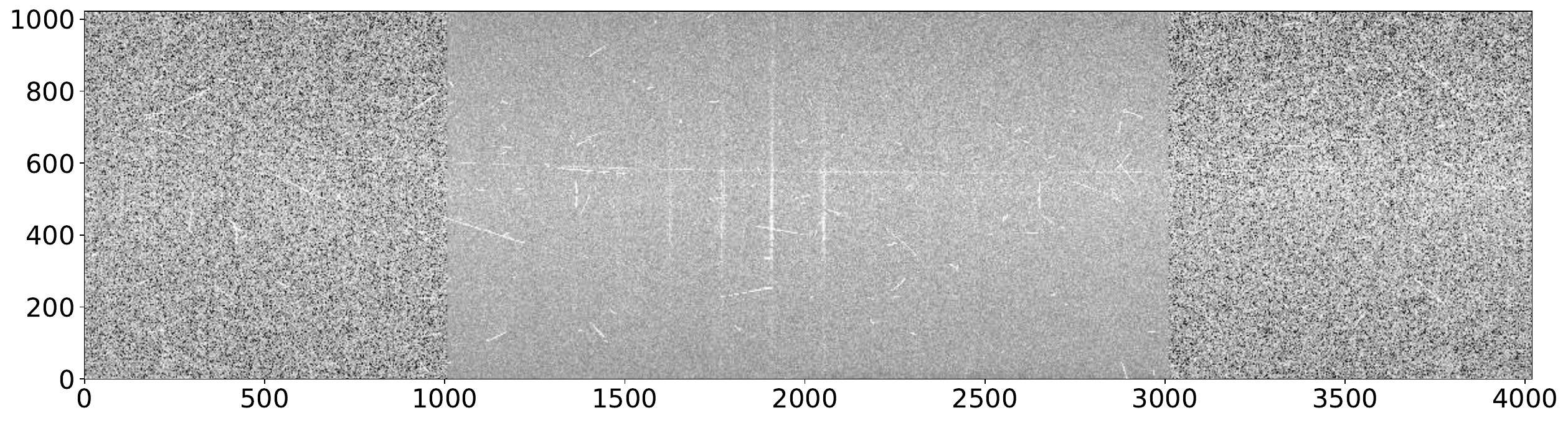} 
      \includegraphics[width=0.69\textwidth]{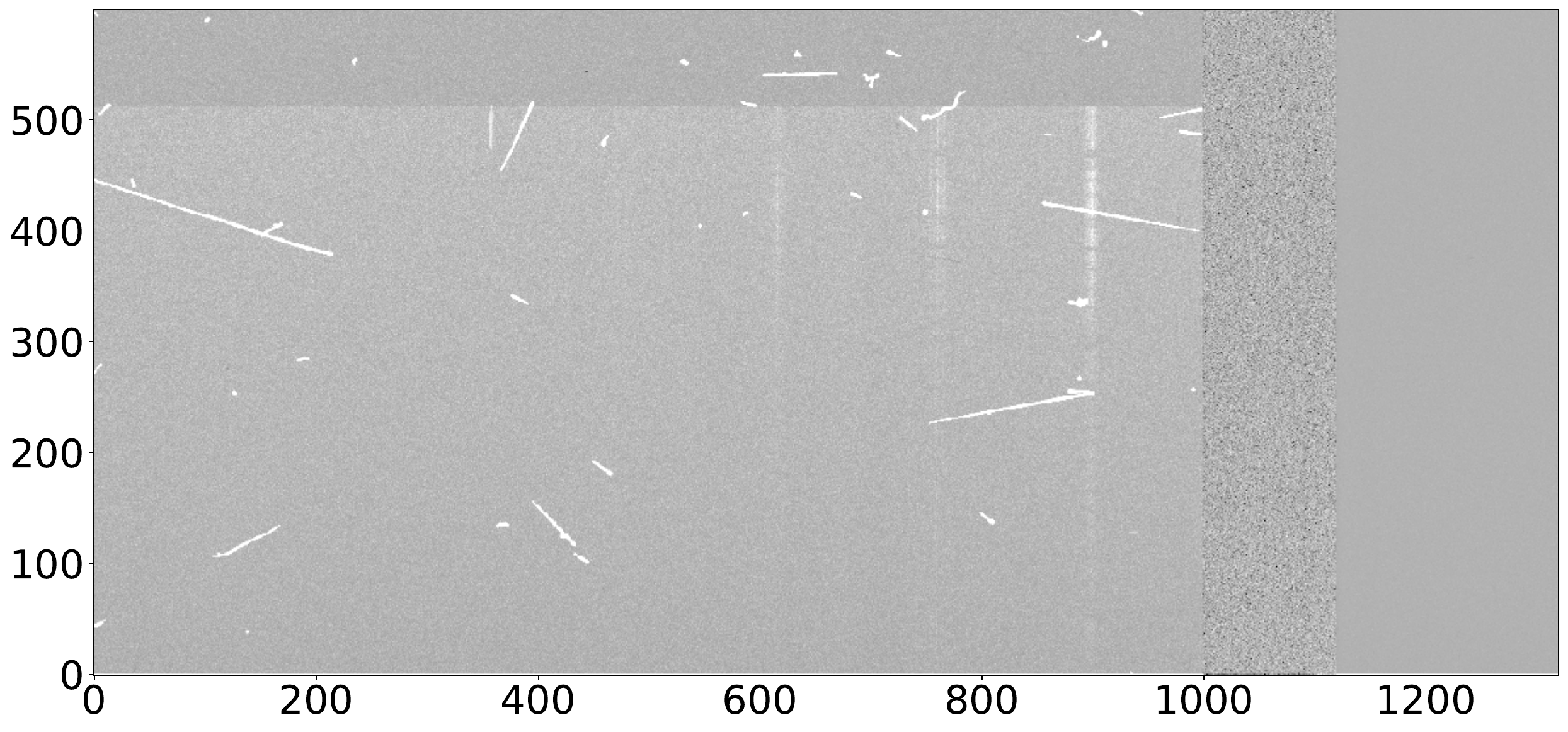}

    \vspace{1em}
    \caption{SIFS-Skipper CCD observation of quasar HB89\,1159$+$123 using the 1/2 FoV ROI readout configuration. {\bf Top:} The 1/2 FoV ROI consists of 50 repeated measurements of 1k $\times$ 2k contiguous physical pixels in the center of each AstroSkipper detector. The top panel shows the physical pixels from the four amplifier quadrants on a single AstroSkipper detector within the SIFS FoV and excludes overscan regions. The x-direction represents the SIFS spatial dimension over the SIFS FoV ($\sim 2$k pixels per amplifier, $\sim 4$k pixels for the detector frame) and the y-direction is the spectral dimension (512 pixels per amplifier, $\sim 1$k for the detector frame). This configuration was used to observe QSOs with sub-electron readout noise covering a focal plane region of 4k $\times$ 2k pixels. {\bf Bottom:} The multi-sample and overscan regions for the bottom left amplifier quadrant from the 1/2 FoV ROI shown in the top image. The multi-sample active area (1k pixels in the x-direction) is followed by a 120-pixel-wide single-sample serial overscan and a 200-pixel-wide multi-sample serial overscan. The parallel overscan is 88 pixels wide in the row direction. These overscan sections are used in image reduction.}
    \label{fig:FOV_ROI}
\end{figure}

\begin{figure}[t!]
    \centering
    \includegraphics[width=0.55\textwidth]{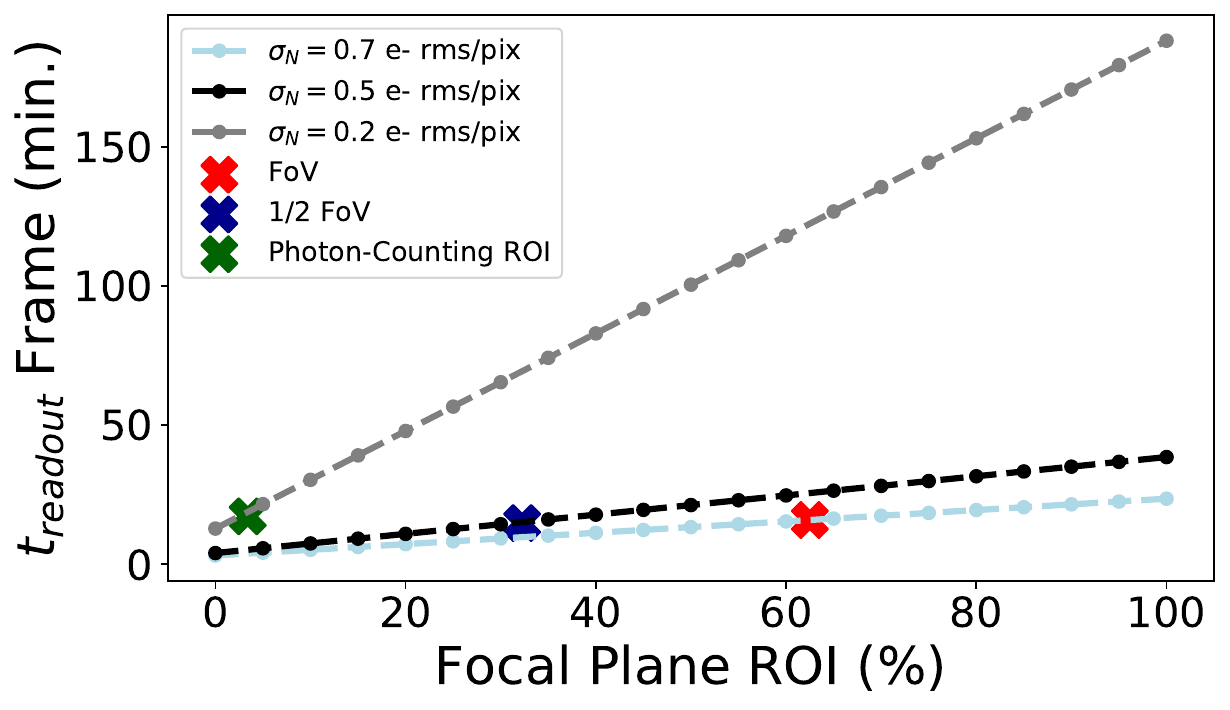}
    \vspace{1em}
    \caption{Readout time for several ROI configurations. The plot gives the expected readout time given the target noise level (sub-electron and photon counting) for different percentages of the detector readout with the target noise. We note that the rest of the FoV is read out with single-sample noise.}
    \label{fig:times}
\end{figure}

The Skipper CCD focal plane spans an area of 6k $\times$ 4k pixels (Figure \ref{fig:focal_plane}); however, the active area where the SIFS spectra are dispersed onto the focal plane is only 4k $\times$ 4k pixels (the ``field of view'', FoV). Figure \ref{fig:FOV} shows the complete focal plane array (6k $\times$ 4k) formed by the four Skipper CCDs; the illuminated fiber bundles correspond to the signal from HD101452, a bright ($V=8$) CALSPEC  standard star \cite{Bohlin_2014}. The bright fiber bundles span the SIFS FoV (pixel $\sim$1000 to $\sim$5000 in the spatial direction). The readout strategy takes advantage of ROIs to readout different regions of the FoV with different readout noise configurations suitable for the astronomical targets. Nominally, the ROIs are defined per detector amplifier quadrant and span 512 physical pixels in the wavelength direction (4k pixels for the whole detector array) and the spatial direction is left as a configurable parameter. The combination of the ROI size and the number of Skipper samples that are collected set the readout time and the readout noise in the ROI (Figure \ref{fig:times}).

The observation sequence begins with a series of calibration products. We take $\sim 5$ flat-field frames illuminating the instrument with a quartz lamp. The instrument focus is set to zero to produce a uniform image and avoid localized imperfections. We then take a sequence of eight masked flat-fields to map the crosstalk between fibers by fitting the amplitude of Gaussian profiles to the spatial cuts across the detector. Flat-field data taking is followed by a series of arc-lamp calibrations using mercury (Hg), argon (Ar), and neon (Ne) lamps to derive the wavelength solution for our spectroscopic data. The full calibration sequence is performed again at the midpoint and end of the night to account for shifts in the fiber positions. The LTA readout electronics do not perform clock shaping, and we found a strong dependence between the amplitude of the horizontal clock voltage swings and the CIC rate \cite{Villalpando_2024}. To maintain high full-well capacity for the calibration images (where the expected signals are $>20{,}000$ e$^-$), the detector clock voltages are set to a high voltage configuration. The CIC rate for these high voltages ($\sim3$ e$^-$/pixel/frame) is negligible compared to the signal. When observing science targets, we switch to an intermediate set of clock voltages with a CIC rate of $< 1$ e$^-$/pixel/frame, which contributes $<10 \%$ of the total noise for sources contributing tens of electrons.

\subsection{Astronomical Targets}

\textit{Quasars:} We identified two quasars, HB89\,1159$+$123 ($z=3.522$) and QSO\,J1621$–$0042 ($z=3.711$) that were observed with the  X-shooter spectrograph on the European Southern Observatory Very Large Telescope as part of the XQ-100 legacy survey \cite{refId0}.  XQ-100 provides high-quality echelle spectra of 100 quasars at redshifts $z \simeq 3.5-4.5$ with high signal-to-noise (median $S/N$ = 30). These high-quality spectra can be compared to the SIFS-Skipper CCD data. We also acquired spectra of these QSOs with the original SIFS Teledyne e2v detector, which serves as another baseline for comparison. 

\noindent \textit{Galaxy Clusters:} We identified moderate-redshift (z $
\sim 1-2$) galaxy clusters intended for the ALP dark matter study. We acquired spectroscopic data with Skipper CCDs from two galaxy clusters: CL\,J1001$+$0220 ($z = 2.5$) and SPT-CL\,J2040$-$4451 ($z=1.48$). These observations are paired with observations of CALSPEC \cite{Bohlin_2014} standard spectroscopic calibration stars (HD074000, HD60753, HD106252, HD101452, HD200654) for flux calibration purposes. Integration times for calibration star exposures are about 5 minutes due to the high signal counts ($> 10,000$ e$^-$) and are read out in single-sample configuration. Previously, we acquired spectroscopic data from these galaxy clusters using the original SIFS e2v detector to compare future science results between both data sets (Section \ref{sec:sci}).     

\noindent \textit{Emission Line Galaxy:} We observed an emission line galaxy (ELG) at $z=0.3239$ that was also observed by DESI as part of the DESI Early Data Release \cite{DESI2023}. The  O\,{\sc iii} emission line is well placed on the the SIFS FoV at $\lambda = 6628.8$\,Å. This observation shows the capability of the Skipper CCDs to resolve faint spectral features and provides a validation of this technology for future spectroscopic surveys (e.g., DESI-2 \cite{schlegel2022spectroscopic}).

\noindent \textit{Ultra-faint Dwarf Galaxy Candide Member Star:} We acquired spectroscopic data associated with a candidate member star in the periphery of the Bo\"{o}tes II ultra-faint dwarf galaxy (UFD) identified by Pan et al.~(2024) \citep{pan2024stellar}. The candidate member star is relatively faint (Gaia $G \sim 19$\,mag), and we plan to measure the radial velocity of this star to determine whether it is a member of  Bo\"{o}tes II (Section \ref{sec:sci}).

\subsection{Sub-Electron Noise Regions of Interest}
We define a ``full FoV'' ROI that covers the entire 4k $\times$ 4k active region of SIFS focal plane array. The full FoV ROI is configured per amplifier with a contiguous active area of 512 $\times$ 2k pixels. The full FoV ROI is read out with $N_{\rm samp}=30$, resulting in a readout noise of $\sim 0.7$\,e$^-$rms/pixel while the the rest of the amplifier region (512 $\times$ 1k pixels outside the FoV) is read with $N_{\rm samp} = 1$. We use this configuration for the galaxy cluster observations and acquired five science exposures with exposure times varying from 900 to 1200 seconds. 

We also define a smaller ``1/2 FoV'' ROI covering a 4k $\times$ 2k region of the focal plane array (excluding the parallel and serial ovserscans).  The 1/2 FoV ROI consists of 512 $\times$ 1k pixels per amplifier where we achieve a noise of $\sim 0.5$\,e$^-$rms/pixel with $N_{\rm samp} = 50$ measurements per pixel. The 1/2 FoV  ROI enables signal to be collected from several of the brightest fiber bundles, and this readout configuration was used to acquire data from the two QSOs, the ELG, and the Bo\"{o}tes II candidate member star. In the top panel of Figure \ref{fig:FOV_ROI}, we show an observation of HB89\,1159$+$123 using the 1/2 FoV ROI configuration; the ROI spans the entire detector in the spectral direction and $\sim$2k pixels in the spatial direction. In the bottom panel of Figure \ref{fig:FOV_ROI}, we show one amplifier segment of a science image to highlight its various components. Following the ROI, there is a single-sample serial overscan that extends for 120 pixels and a multi-sample serial overscan that extends the remaining 200 pixels. Additionally, the parallel overscan begins at pixel 512 and extends to pixel 600 in the wavelength (row) direction maintaining the same number of samples per pixel. From Figure \ref{fig:times}, we note that the readout time is $\sim 15$ min for both FoV ROI configurations.

\begin{figure} [t!]
   \begin{center}
   \begin{tabular}{l l} 
    \includegraphics[height=5.5cm]{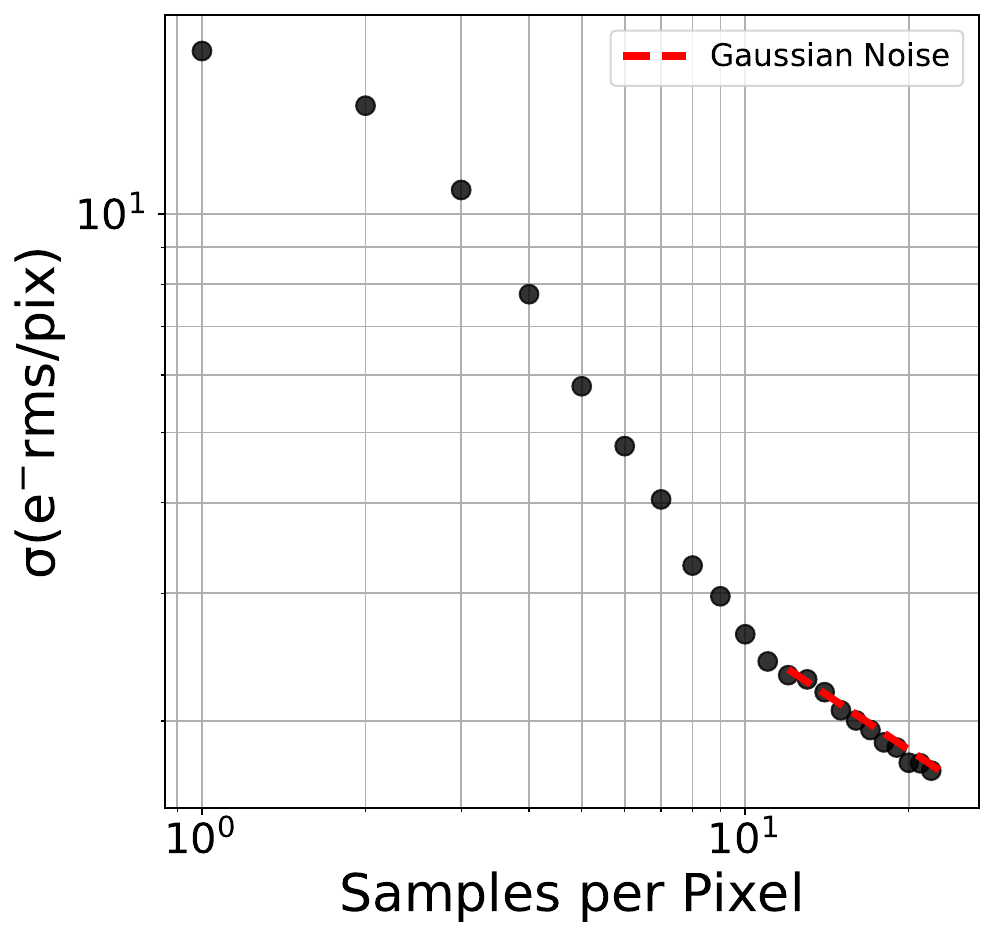} 
   \includegraphics[height=5.5cm]{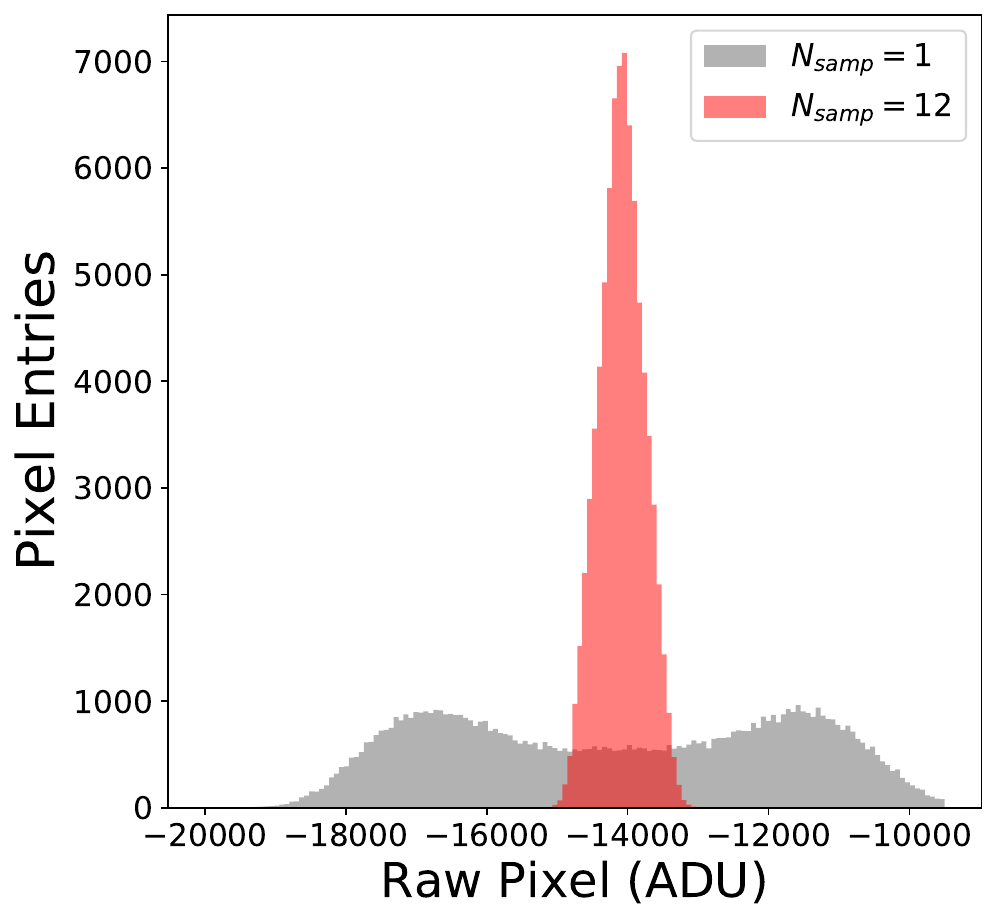}    
   \end{tabular}
   \end{center}
   \caption{\label{fig:readout_noise} 
   \textbf{Left:} Readout noise from one of the  detectors affected by correlated noise as a function of the number of samples per pixel ($N_{\rm samp}$). The single-sample readout noise is $\sim 17$\,e$^-$\,rms/pixel but rapidly decreases with increasing $N_{\rm samp}$, reaching $\sim 4.6$ e$^-$ rms/pixel at $N_{\rm samp} = 7$. The theoretical Gaussian noise behavior (i.e., $\sigma_N \propto 1/\sqrt{N_{\rm samp}}$) is recovered for $N_{\rm samp} > 12$. These detectors are able to count photons at $N_{\rm samp}=250$ similarly to the two unaffected devices. \textbf{Right:} The pixel distribution from the multi-sample serial overscan. For small $N_{\rm samp}$, the pixel distribution exhibits a ``double-Gaussian'' feature, which suggests a temporal correlation between mixed sources of noise. For $N_{\rm samp} \gtrsim 12$, a single-Gaussian noise distribution is recovered.}
\end{figure} 

\begin{figure} [t]
   \begin{center}
   \includegraphics[height=4.5cm]{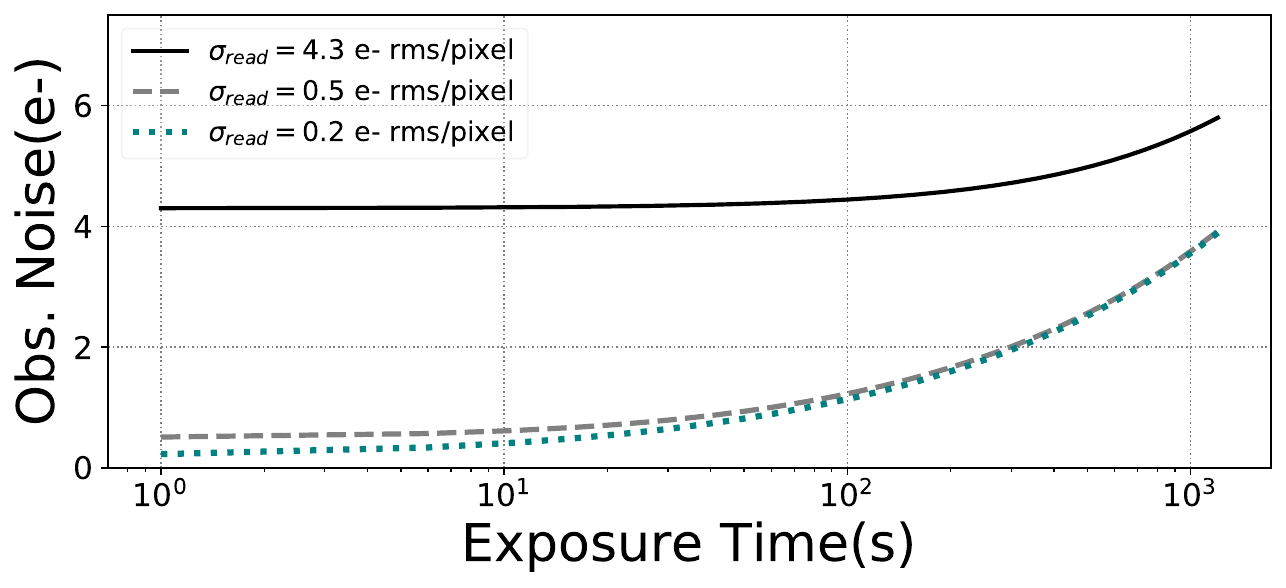}    
   \end{center}
   \caption{\label{fig:noise} 
    The total observation noise as a function of the exposure time. The total observation noise includes the background shot noise. In the configuration shown here, the background counts are dominated by the shutter-open dark rate.  We note that for short exposure times (e.g., the $\sim 60$\,s exposure time used for some photon-counting readout configurations), the shot noise is subdominant to the readout noise.}
\end{figure} 

\subsection{Photon-Counting Region of Interest}
As a demonstration of the first on-sky photon-counting spectroscopic data collected with a Skipper CCD, we defined a ``photon-counting'' ROI spanning 220 pixels in the spatial direction (110 pixels per amplifier), which we sample 250 times to achieve a readout noise of $\sim 0.22$\,e$^-$rms/pix. The photon-counting ROI was used to acquire spectra from HB89\,1159$+$123 and QSO\,J1621$-$0042. The blaze angle was chosen to center the Lyman-alpha (Ly-$\alpha$) line (${\sim} 5497$ Å and ${\sim} 5726$ Å for the two QSOs, respectively) on a single detector. The readout time for this configuration is 17.2 min (Figure \ref{fig:times}). We took one 120\,s science exposure from HB89\,1159$+$123 in order to validate noise and photon-counting performance, followed by two 60\,s exposures for each QSO.   

\section{RESULTS}
\label{sec:results}  
\subsection{System Performance}
To the best of our knowledge, these observations represent the first-ever on-sky data collected  with Skipper CCDs. They demonstrated the capability of these detectors to achieve ultra-low noise and detect weak signals (e.g., resolve individual photons) under real instrument and observing scenarios.
This successful demonstration of Skipper CCDs on the sky relies not only on the demonstrated performance of the detectors \citep{Villalpando_2024}, but also on the integrated performance of the focal plane electronics, instrument, and telescope \citep{Bonati:2024}. We note two minor system issues that affected these first observations.

When integrating the Skipper CCD focal plane, we found that one pair of detectors exhibited unusually high correlated single-sample noise ($\sim 17$ e$^-$ rms/pixel) compared to the other pair of devices ($< 4.6$ e$^-$ rms/pixel) \cite{Bonati:2024}. Since these two detectors share clocks and biases, we attribute the source of this noise to an electronic coupling between these detectors. While we were unable to identify the precise source of this noise in the lab, we showed that this correlated noise quickly reaches nominal noise performance ($\sim 4.6$\,e$^-$\,rms/pixel) with $N_{\rm samp}=7$. Furthermore, as shown in Figure \ref{fig:readout_noise}, the Gaussian noise scaling behavior with $N_{\rm samp}$ is observed to recovered for $N_{\rm samp}>12$. We measure a readout noise of $\sim 0.72$\,e$^-$\,rms/pixel and $\sim 0.22$ e$^-$ rms/pixel (enabling photon-counting) after $N_{\rm samp}=50$ and $N_{\rm samp}=250$, respectively. This multi-sample noise performance is comparable to the two unaffected devices.

During the first engineering night, we measured a shutter-open dark rate of $\mathcal{O}(10^{-1}$)\,e$^-$/pixel/s, which was attributed to stray light within the SIFS enclosure. We identified and mitigated several light sources contaminating the instrument's optical path. Science observations proceeded with an improved background rate $\sim 1.19 \times 10^{-2}$ e$^-$/pixel/s. While this remains sub-optimal,  the stray light shot noise contribution can be kept to $\lesssim 2.5$\,e$^-$\,rms/pixel if the exposure time is limited to $\lesssim 600$\,s (Figure \ref{fig:noise}). We are exploring options to reduce the background rate to levels measured during laboratory testing ($\sim 2 \times 10^{-4}$ e$^-$/pixel/s) \cite{Villalpando_2024}.   

\begin{figure}[t!]
\centering 
 \includegraphics[height=9.5cm]{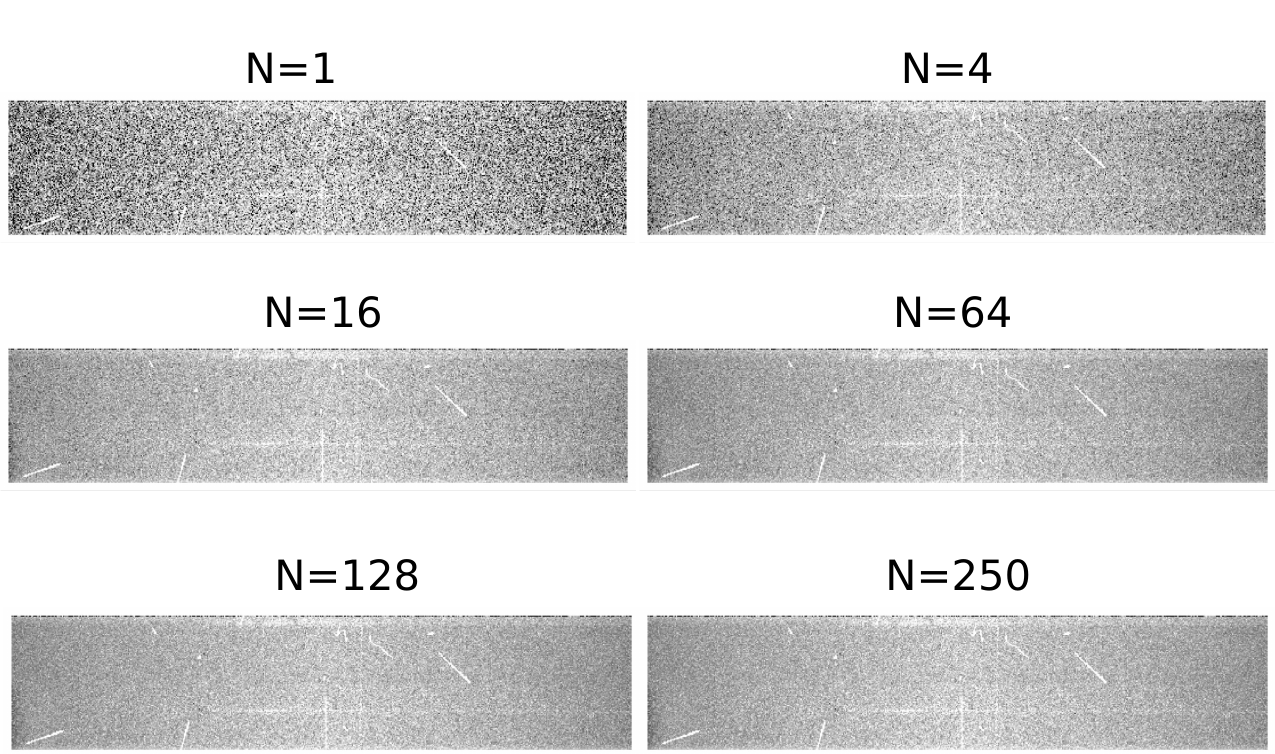} 
\caption{\label{fig:image_section} Transition between single-sample to photon-counting readout noise level for a data frame containing the Ly-$\alpha$ line from HB89\,1159$+$123. Each sequence shows an increasing number of non-destructive pixel measurements and decreasing readout noise. The faint spectral trace becomes increasingly visible as the readout noise decreases. Note that the spatial and spectral axes have been flipped relative to Figures \ref{fig:FOV} and \ref{fig:FOV_ROI}.}
\end{figure}

\begin{figure}[t!]
\centering 
  \includegraphics[height=20.5cm]{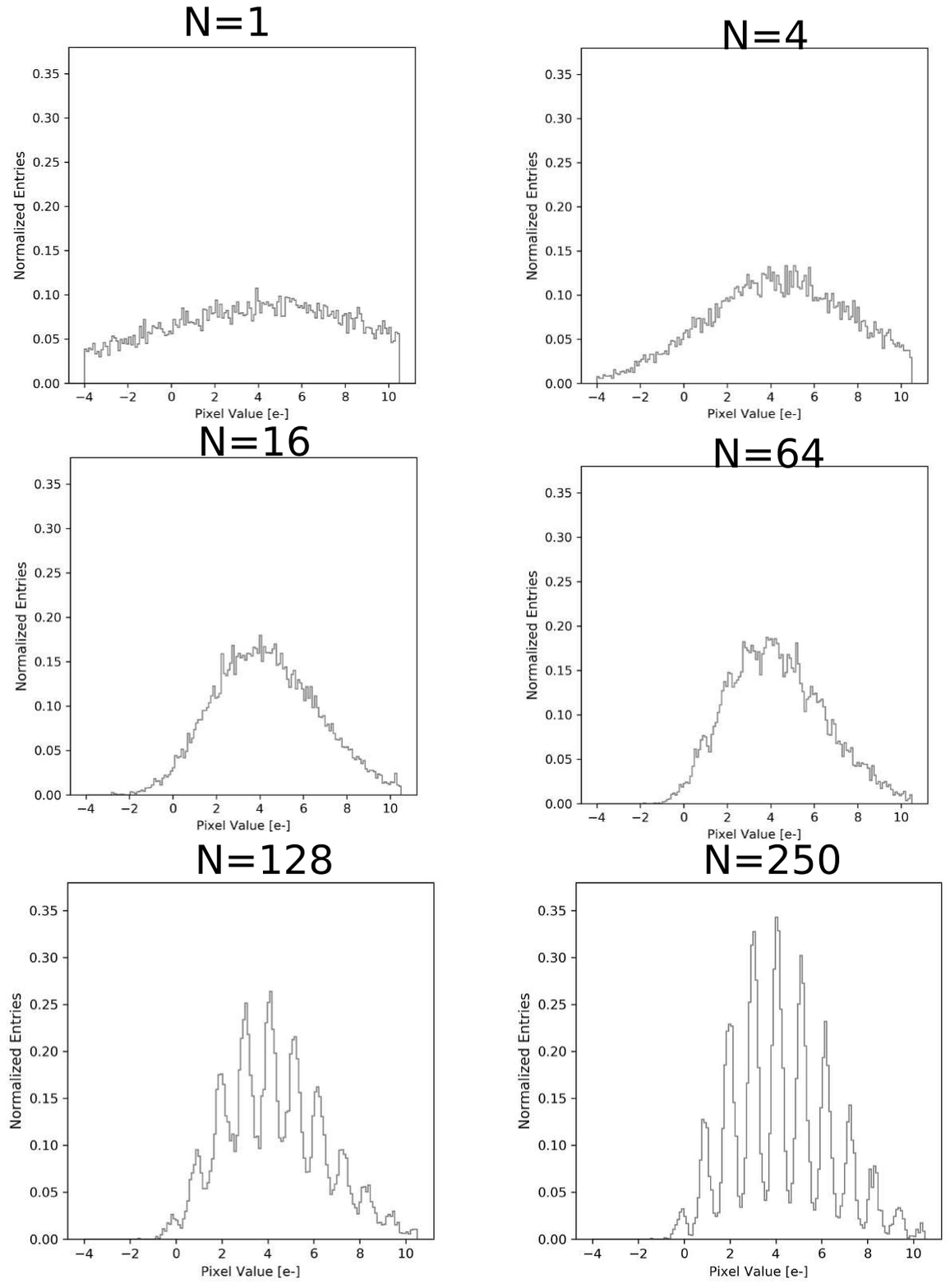}
  \caption{\label{fig:photon} Photon-counting demonstration on the sky with the SIFS Skipper CCD focal plane. Photon resolution is achieved at 128 and 250 measurements per pixel over the ROI shown in Figure \ref{fig:image_section}. }   
\end{figure}

\clearpage

\begin{figure}[t!]
  \centering 
  \includegraphics[height=13cm]{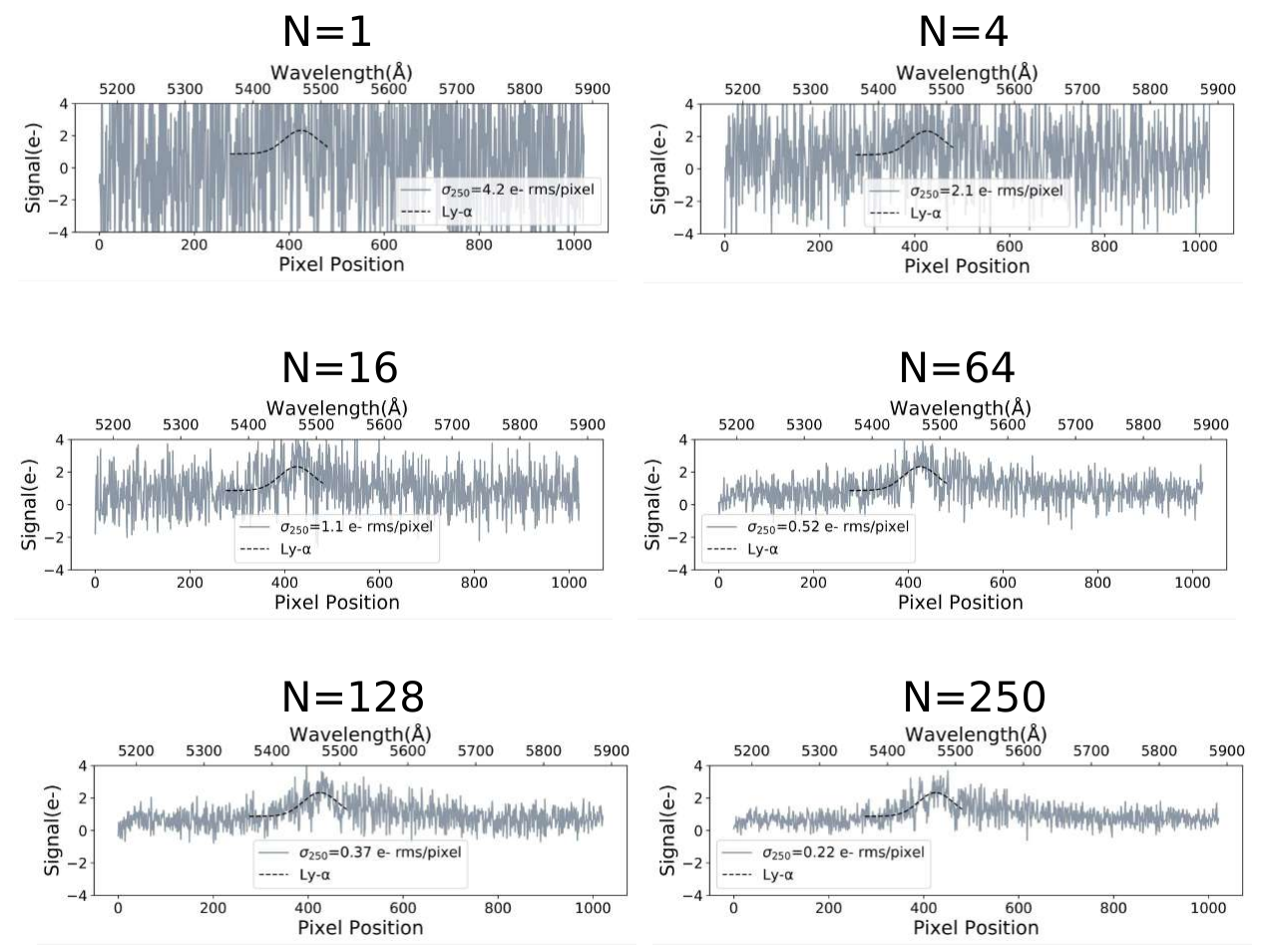}
\caption{\label{fig:qso_spectra_photon} Noise reduction sequence illustrating the Ly-$\alpha$ line from the data frame in Figure \ref{fig:image_section}. Of particular interest for this demonstration is the ability to resolve features of interest in the spectra. The  Ly-$\alpha$ line is completely hidden in the readout noise for low number of samples. As the noise decreases down to the photon-counting regime, one is able to identify and precisely fit the Ly-$\alpha$ line.}   
\end{figure}

\subsection{Single-Photon Counting}
Single-photon counting was achieved for the first time with a Skipper CCD on the sky. We attained a readout noise of $\sim 0.22$\,e$^-$rms/pixel over $\sim$5.4$\%$ of the SIFS FoV (Section \ref{sec:obs}), enabling photon-counting in all detectors. Figures \ref{fig:image_section}, \ref{fig:photon}, and \ref{fig:qso_spectra_photon} illustrate the ultra-low noise capabilities of the Skipper CCD through a sequence of readout noise configurations, ranging from $N_{\rm samp} =1$ ($\sim 4.6$ e$^-$ rms/pixel) to $N_{\rm samp}=250$ ($\sim 0.22$\,e$^-$\,rms/pixel). The frames in Figure \ref{fig:image_section} represent a 60 second exposure of the photon-counting ROI from one of the detectors, containing a bright fiber bundle which corresponds to the Ly-$\alpha$ line in HB89\,1159$+$123. As the number of samples increases and the readout noise deceases, the line becomes increasingly visible. The reduction in the readout noise enables photon/electron counting (Figure \ref{fig:photon}) and the Ly-$\alpha$ feature becomes detectable above the noise (Figure \ref{fig:qso_spectra_photon}). This is a powerful demonstration of the ultra-low noise capabilities the Skipper CCD provides. With the short exposure, we purposefully reduce the signal-level from the QSO while reducing the background contribution to the noise ($< 1$ e$^-$)  to demonstrate the ability for the Skipper CCD to resolve spectral features contributing a few electrons.

\begin{figure}
\centering 
 \includegraphics[clip,width=1.0\columnwidth]{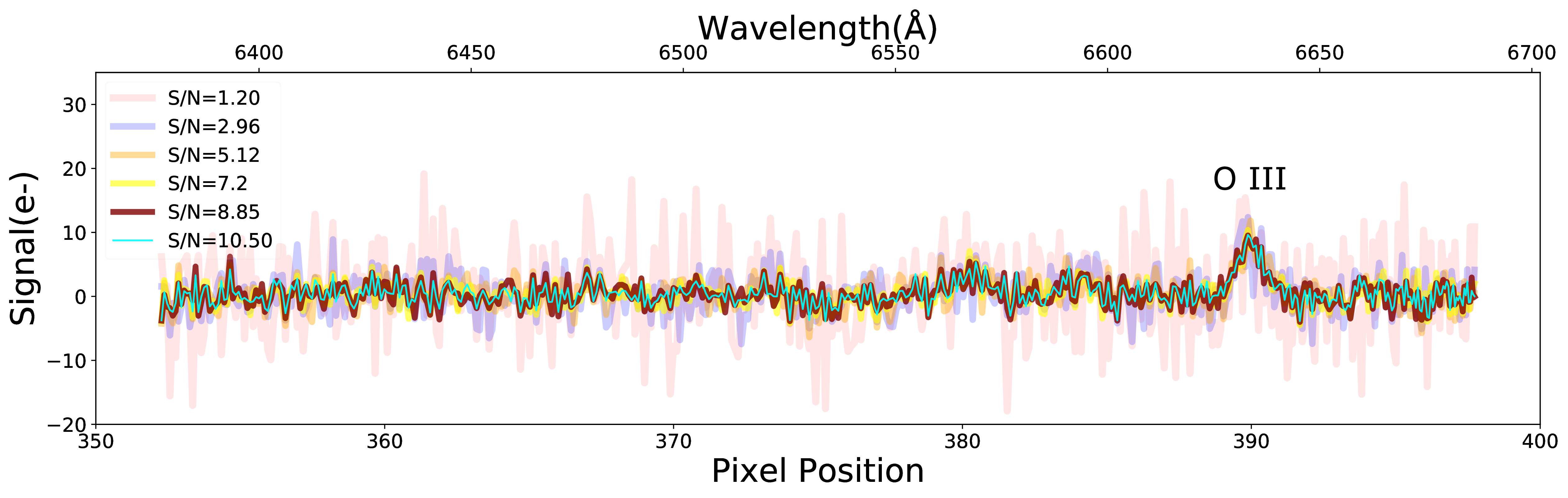}\\
 \includegraphics[clip,width=0.55\columnwidth]{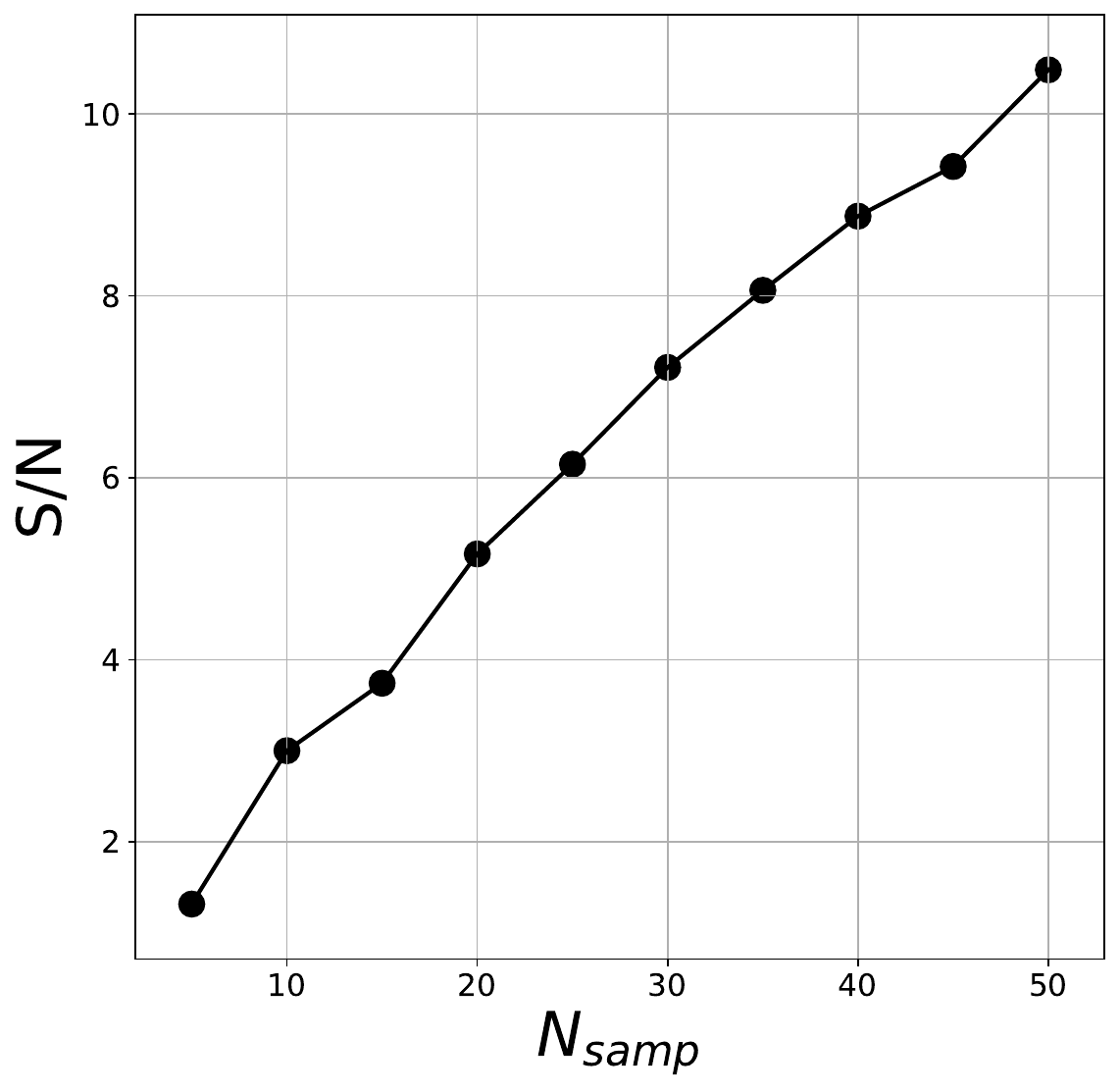}\\
\caption{\label{fig:elg} \textbf{Top}: Spectrum from a 600\,s SIFS-Skipper CCD observation  of an ELG identified in the DESI Early Data Release. The  O\,{\sc iii} emission line is detected at $\lambda \sim 6628.8$ Å with different signal-to-noise ratios depending on the achieved readout noise (colored lines). \textbf{Bottom}: The signal-to-noise ratio as a function of $N_{\rm samp}$. In the readout-noise-dominated regime, it is possible to increase the SIFS signal-to-noise from ${\sim} 1.2$ to  ${\sim} 10.5$ by achieving sub-electron readout noise levels. Note that this observations was taken by one of the Skipper CCDs that was affected by correlated noise (Section \ref{sec:comissioning}).}
\end{figure}

\begin{figure}
\centering 
 \includegraphics[clip,width=1.0\columnwidth]{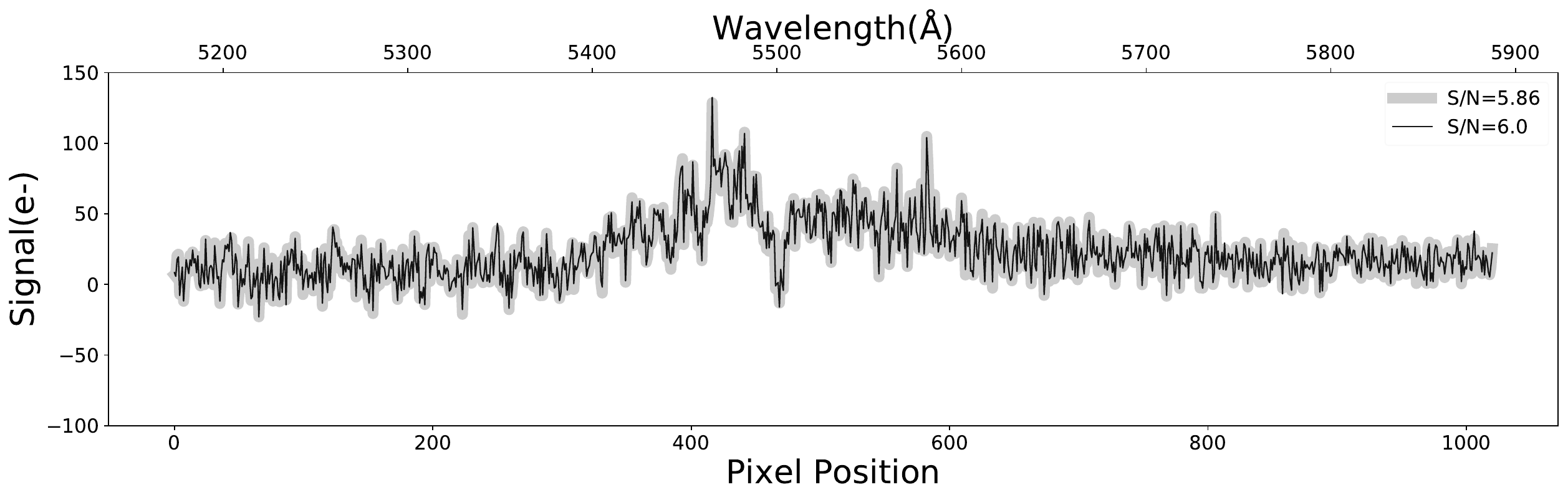}\\
 \includegraphics[clip,width=0.55\columnwidth]{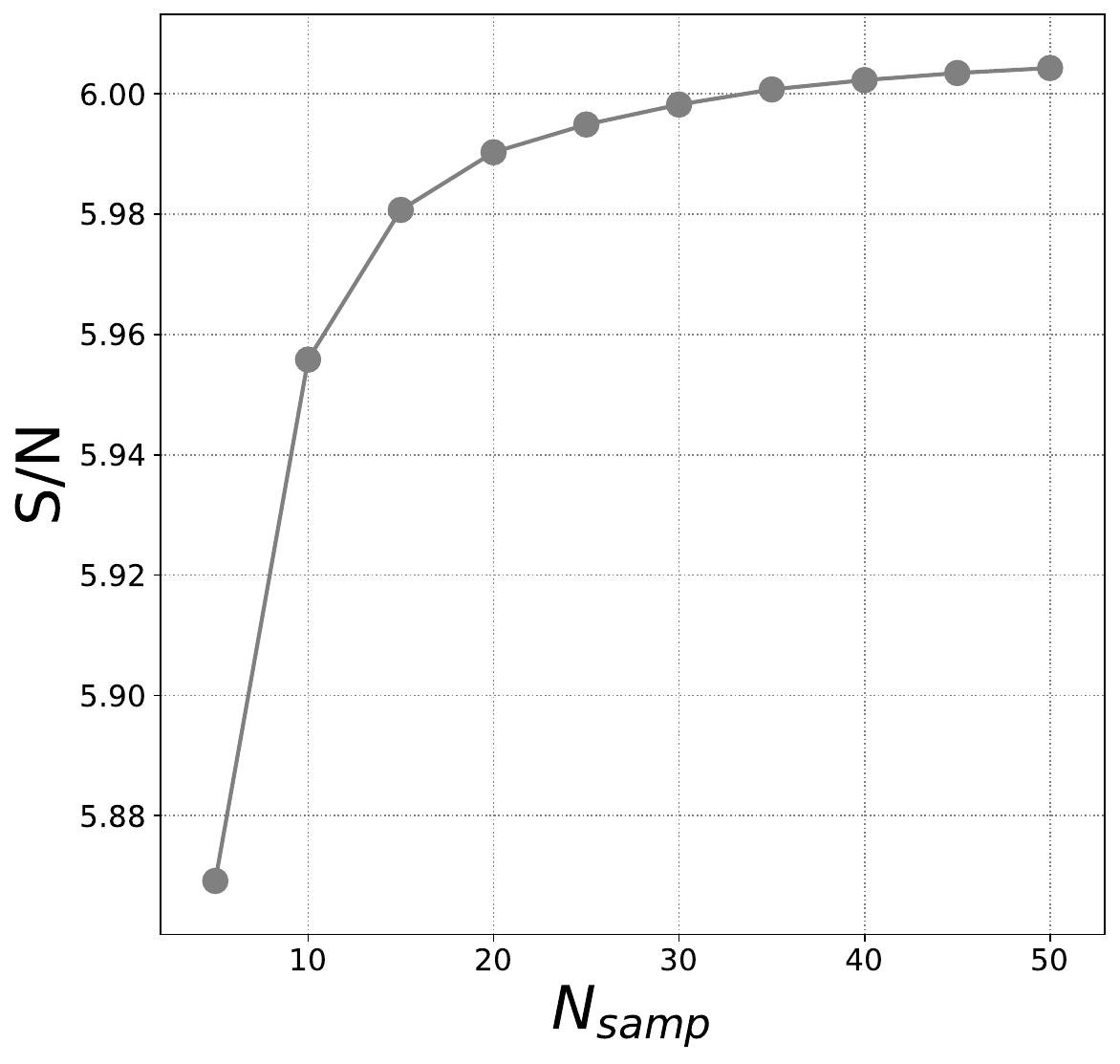}\\
 \caption{\label{fig:high_signal} \textbf{Top}: High signal level (${\sim} 70$ e$^-$) spectrum around the Ly-$\alpha$ feature from a SIFS-Skipper CCD observation of QSO HB89\,1159$+$123. \textbf{Bottom}: Signal-to-noise as a function of $N_{\rm samp}$ for the high signal QSO observation. The signal-to-noise improvement is minimal. The source and background contribution to the total noise is $\sim 10$ e$^-$, which dominates the noise budget in this observation.}
\end{figure}

\subsection{Signal-to-Noise Improvement}
The signal-to-noise ratio is a measure of the ability to distinguish a signal from background noise.
For faint sources in low-background observations, the readout noise can dominate the total noise in the observation, i.e., 
\begin{align}
\sigma_{obs.}= \sqrt{\sigma_{shot}^2 + \sigma_{read}^2}
\label{eqn:total_noise}
\end{align}
where $\sigma_{shot}^2$ represents the Poisson shot noise and $\sigma_{read}$ is the detector's readout noise. The Poisson shot noise is calculated from the electron count rate (e$^-$/pixel/frame) from all sources, including CIC and the detector's ``dark rate'', which includes external light contamination leaking into the system. The Poisson shot noise, in contrast to the readout noise, increases with the exposure time and becomes the dominant component of the total observation noise when signal and/or background levels are high. In our current system, the background rate is sub-optimal due to issues with external light leakage. By reducing the exposure time, we can reduce the shot noise contribution from the signal and backgrounds in order to explore signal-to-noise improvements from the Skipper CCDs in the readout-noise-dominated regime. Figure \ref{fig:noise} illustrates this interplay between exposure time (count level) and the readout noise. In exposures with few counts, reducing the readout noise can lead to significant gains in signal-to-noise. 

Figure \ref{fig:elg} shows the  O\,{\sc iii} feature from the ELG for a 600 second exposure from one of the Skipper CCDs affected by correlated noise. We show that as the readout noise decreases from $\sim 6$ e$^-$ rms/pixel with $N_{\rm samp}=5$ to $\sim 0.7$ e$^-$ rms/pixel after $N_{\rm samp}=50$, the signal-to-noise ratio increases up to $10.5$, allowing for the O\,{\sc iii} emission line to be detected. In contrast, Figure \ref{fig:high_signal} illustrates the Ly-$\alpha$ line from HB89\,1159$+$123 with higher signal ($\sim 70$ e$^-$ on average) and a total of $\sim 10$ e$^-$ contribution from the background. In this instance, the signal-to-noise stays roughly consistent because the low readout noise is subdominant (e.g., $\sim 0.5$ e$^-$ rms/pixel after $N_{\rm samp}=50$).

\section{SCIENTIFIC APPLICATIONS}
\label{sec:sci}
We identified two scientific applications for the SIFS Skipper CCD science verification data: (1) we will use observations from the galaxy clusters (CL\,J1001$+$0220 and SPT-CL\,J2040$-$4451) to place new constraints on the mass of ALP dark matter, and (2) we will use spectra from the candidate member star of the Bo\"{o}tes II UFD to further study the spatial distribution and dark matter halo of this system.  We expect that the low readout noise from the Skipper CCD focal plane will improve the results from these two scientific cases. The low readout noise allows faint lines to become visible above the background, which is crucial when deriving constraints on the ALP-photon coupling. Additionally, the improved signal-to-noise for the UFD candidate   member star enables a more precise measurement of its radial velocity.   

\subsection{ALP Dark Matter}
\label{sec:alps}
\begin{figure}
    \centering
    \includegraphics[width=0.7\textwidth]{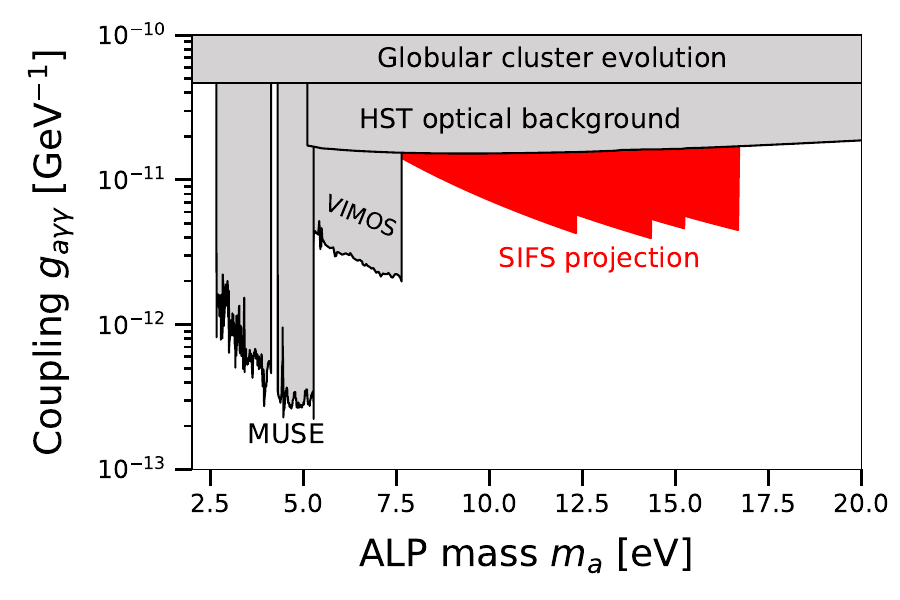}
    \caption{Collection of selected constraints on the ALP mass/coupling parameter space. Previous results (gray) include integral-field spectroscopic data from MUSE ~\cite{REGIS2021136075} and VIMOS~\cite{PhysRevD.75.105018}, HST optical background~\cite{2023PhRvD.107h3032C}, and stellar evolution~\cite{Dolan:2022kul}. Projected constraints from observation of galaxy clusters with SIFS (described in Sec.~\ref{sec:alps}) are shown in red, demonstrating the advantages of low-noise detectors, as well as observations of higher-redshift clusters. }
    \label{fig:alp_constraints}
\end{figure}

ALPs are a generalization of the Quantum chromodynamics (QCD) axion, initially proposed to solve the strong charge-parity (CP) problem \cite{PhysRevLett.38.1440, PhysRevD.16.1791}. ALPs are also considered a viable dark matter candidate, and many efforts to search for them exploit their coupling to photons. From an astrophysics perspective, these searches aim to detect photon signals resulting from the radiative decay of ALP dark matter and the conversion of ALPs into photons in the presence of magnetic fields. For ALPs with masses in the electron-volt (eV) scale, the monochromatic photon emission from ALP decay falls within the optical and infrared wavelength range \cite{PhysRevD.75.105018, REGIS2021136075}. Under the proposal SO2024A-011 (``Searching for decaying axion like particles in $z > 1.1$ galaxy clusters with SOAR/SIFS''), we have taken five science exposures of Sunyaev-Zeldovich-selected galaxy clusters (e.g., \cite{Huang_2020}) with the Skipper CCD focal plane prototype (more observations are planned for early July 2024). These include two 1200 second exposures of the cluster CL\,J1001$+$0220 ($z \sim 2.51$, ~\cite{2016ApJ...828...56W}) and three 900 second exposures of SPT-CL\,J2040$-$4451 ($z \sim 1.48$, ~\cite{2014ApJ...794...12B}). For both clusters, we read out the full FoV with $N_{\rm samp} = 30$, providing ${\sim}0.7\text{\,e}^-$\,rms/pixel readout noise on the two detectors not affected by the correlated noise while maintaining a manageable level of cosmic-ray contamination during read out. We show in Fig.~\ref{fig:alp_constraints} the expected sensitivity of these SIFS observations, compared to previous constraints from MUSE, VIMOS, and HST. (VIMOS was --- and MUSE is --- an integral-field spectrograph on the Very Large Telescope (VLT).) Despite the smaller diameter of SOAR compared to the VLT (4.1 m vs.\ 8.2 m), and the smaller FoV of SIFS compared to MUSE/VIMOS (0.03 arcmin$^2$ vs.\ 1 arcmin$^2$), we anticipate SIFS will deliver similar constraints on the ALP-photon coupling $g_{a\gamma\gamma}$ in a higher-mass range than has previously been probed. This is a combination of the low readout noise of the Skipper CCDs (allowing faint lines to become visible above the sky background) and observing higher-redshift galaxy clusters (allowing photons from higher-mass --- and thus faster-decaying --- ALPs to redshift into the SIFS bandpass).

\subsection{Bo\"{o}tes II Candidate Member Star}
Pan et al.\ (2024) \cite{pan2024stellar} conducted a photometric analysis using metallicity sensitive DECam $u$-band imaging to identify candidate member stars in the outskirts of three Milky Way UFDs: Bo\"{o}tes I, Bo\"{o}tes II, and Segue~1. The study identify UFD candidate member stars from derived photometric metallicities, and the authors of that work recommended spectroscopic follow-up to confirm the newly-identified candidate member stars. The SIFS-Skipper CCD system is well-suited to measure the the radial velocities of these candidate member stars to unambiguously determine their membership and extend the mapping of the stellar populations of UFDs to large distances. UFDs form within extended dark matter halos and serve as ideal laboratories to study the nature of dark matter at small scales \cite{Walker_2009, Munshi_2019, 10.1093/mnras/stz2887}. Tracing the outer stellar distributions of UFDs is important to probe the relationship between dwarf galaxies and their small dark matter halos. We acquired four 900 second exposures targeting a Bo\"{o}tes II candidate member star previously identified in Pan et al.\ (2024) \cite{pan2024stellar}. We expect to determine the star's radial velocity from the H-$\alpha$ and/or Mg triplet absorption lines.

\section{SUMMARY and OUTLOOK}
\label{sec:end}  

We have presented results from the first-ever on-sky demonstration of  ultra-low noise, photon-counting Skipper CCDs at SIFS on the 4.1-m SOAR Telescope. Astronomy-optimized Skipper CCDs show performance parity with DECam and DESI devices (Table \ref{tab:astroskipper}), demonstrating the viability of Skipper CCDs for measuring faint spectroscopic signals significantly below the noise levels in conventional scientific CCDs. We outlined an observation strategy that exploits ROIs to achieve sub-electron ($\lesssim 0.7$ e$^-$ rms/pixel) and photon-counting ($\sim 0.22$ e$^-$ rms/pixel) readout noise levels in portions of the FoV in $\lesssim 17$ min. 

We acquired spectroscopic data with sub-electron readout noise levels from two QSOs at redshift $z \sim 3.5$ (HB89\,1159$+$123 and QSO\,J1621$–$0042), two moderate-redshift galaxy clusters (CL\,J1001$+$0220 and SPT-CL\,J2040$-$4451), an ELG at $z = 0.3239$, and a candidate member star of the Bo\"{o}tes II UFD. In addition to performing the first demonstration of the Skipper CCD technology on the sky, we expect to derive scientific results from these data products. As a demonstration of the ultra-low noise of the Skipper CCD, we acquired spectra from HB89\,1159$+$123 over $\sim 5.4 \%$ of the FoV with a readout noise of $\sim 0.22$\,e$^-$\,rms/pixel. We resolved individual photo-electrons and demonstrated the ability to detect the Ly-$\alpha$ emission line. Furthermore, we show that when the observation's total noise is dominated by the readout noise, it is possible to significantly increase the SIFS signal-to-noise ratio by lowering the readout noise. This demonstrates the potential to increase the sensitivity of future observations of faint sources in the readout-noise-dominated regime.

Achieving sub-electron and photon-counting readout noise with Skipper CCDs comes at the cost of long readout times. To address this, we implemented ROIs in our observations. However, more promising alternatives to minimize readout times include improved detector architectures incorporating the Skipper floating gate amplifier. One such innovation is the Multi-Amplifier Sensing CCD (MAS CCD), which features a series of floating-gate output amplifiers placed along an extended serial register, enabling non-destructive measurements \cite{https://doi.org/10.1002/asna.20230072, 10521851}. 
In contrast to conventional Skipper CCDs, the readout noise from a MAS CCD is given by 
\begin{align}
\sigma= \frac{\sigma_{0}}{\sqrt{N_{s} N_{a}}}
\label{eqn:mas_noise}
\end{align}
\noindent where $N_{a}$ is the number of amplifiers in the serial register and $N_{s}$ is the number of non-destructive measurements taken with each amplifier \cite{https://doi.org/10.1002/asna.20230072, 10521851}. The extra factor of $1/\sqrt{N_{a}}$ in the noise reduction means that  it is possible to achieve ultra-low noise by combining measurements from each amplifier. The improvement in readout time from a MAS CCD compared to a conventional single-amplifier Skipper CCD is ${\sim} 1/N_{a}$. Current work is underway to increase the number of on-chip Skipper amplifiers on MAS devices to achieve fast sub-electron and photon-counting readout noise. A 4k $\times$ 4k MAS CCD would eliminate the need for a detector mosaic configuration with gaps and achieve $\lesssim 1$\,e$^-$\,rms/pixel in $\sim$1 minute of readout time.  

\acknowledgments   

The on-sky data is based on observations obtained at the Southern Astrophysical Research (SOAR) telescope, which is a joint project of the Minist\'{e}rio da Ci\^{e}ncia, Tecnologia e Inova\c{c}\~{o}es (MCTI/LNA) do Brasil, the U.S. National Science Foundation NOIRLab, the University of North Carolina at Chapel Hill (UNC), and Michigan State University (MSU).
The SIFS Skipper CCD observations were taken using time allocated to Brazilian programs on SOAR.
The fully-depleted Skipper CCD was developed at Lawrence Berkeley National Laboratory, as were the designs described in this work.

EMV acknowledges support from the DOE Graduate Instrumentation Research Award and the DOE Office of Science Office of Science Graduate Student Research Award.
The work of AAPM was supported by the U.S. Department of Energy under contract number DE-244AC02-76SF00515. 
This work was partially supported by the Fermilab Laboratory Directed Research and Development program (L2019.011 and L2022.053). 
Support was also provided by NASA APRA award No.~80NSSC22K1411 and a grant from the Heising-Simons Foundation (\#2023-4611).
This manuscript has been supported by NOIRLab, which is managed by the Association of Universities for Research in Astronomy (AURA) under a cooperative agreement with the National Science Foundation.

This manuscript has been authored by the Fermi Research Alliance, LLC, under contract No.~DE-AC02-07CH11359 with the US Department of Energy, Office of Science, Office of High Energy Physics. The United States Government retains and the publisher, by accepting the article for publication, acknowledges that the United States Government retains a non-exclusive, paid-up, irrevocable, worldwide license to publish or reproduce the published form of this manuscript, or allow others to do so, for United States Government purposes.

\renewcommand{\refname}{REFERENCES}

\bibliography{main} 
\bibliographystyle{spiebib} 

\end{document}